\documentclass[twocolumn]{aastex631}
\usepackage{color}
\usepackage[titletoc]{appendix}
\usepackage{amsmath}
\usepackage{amssymb}
\usepackage{mathtools}
\usepackage{upgreek}
\usepackage{comment}
\usepackage{enumitem}
\usepackage{gensymb}
\usepackage{natbib}
\usepackage{graphicx}
\usepackage{bm}
\usepackage{totcount}
\usepackage{multirow}
\usepackage{hyperref}
\usepackage{cleveref}
\usepackage{tabularx}
\usepackage[T1]{fontenc}
\hypersetup{linkcolor=cyan,citecolor=cyan,filecolor=cyan,urlcolor=cyan}

\newtotcounter{citnum} 
\def\oldbibitem{} \let\oldbibitem=\bibitem
\def\bibitem{\stepcounter{citnum}\oldbibitem}

\crefname{subsection}{subsection}{subsections}

\shortauthors{}
\shorttitle{}

\begin{document} 

\title{Shaving the Outskirts of Planetary Systems Probed by Roman via Stellar Flybys}

\author[0009-0003-3584-6698]{Donald Liveoak}
\altaffiliation{Co-first author.}
\affiliation{Department of Physics, University of Michigan, Ann Arbor MI 48109, USA}
\email{dliveoak@umich.edu}

\author[0000-0003-4992-8427]{Tim Hallatt}
\altaffiliation{Co-first author.}
\affil{MIT Kavli Institute for Astrophysics and Space Research, Massachusetts Institute of Technology, Cambridge, MA 02139, USA}

\author[0000-0003-3130-2282]{Sarah C. Millholland}
\affiliation{MIT Kavli Institute for Astrophysics and Space Research, Massachusetts Institute of Technology, Cambridge, MA 02139, USA}
\affiliation{Department of Physics, Massachusetts Institute of Technology, Cambridge, MA 02139, USA}

\begin{abstract}
The recently launched Nancy Grace Roman Space Telescope (\textit{Roman}) will soon begin observations that promise to revolutionize our understanding of exoplanet demographics.
\textit{Roman}'s Galactic Bulge Time-Domain Survey (GBTDS) will observe the Galactic Bulge via transit and microlensing techniques, providing a unique opportunity to study hot Jupiters as well as long-period/free-floating exoplanets in a dense and kinematically hot stellar environment. In this Letter, we quantify how repeated, impulsive stellar flybys truncate the outskirts of planetary systems in the Bulge. We find that flybys result in ejections on $\sim$Gyr timescales for wide orbits (${\gtrsim} 250\ \text{au}$) around low-mass hosts. Provided these wide orbits are populated, these ejections likely contribute to the population of free-floating planets probed by \textit{Roman}. Surviving planets outside of ${\sim} 10$ au are placed on eccentric and misaligned orbits. We show that, though systems suffer many close encounters, the evolution is dominated by the single strongest perturbation; planetary orbits (even those that evade ejection) undergo superdiffusion (i.e. a L\'evy flight). We comment on the implications of our results for the origin of hot Jupiters amenable to $\textit{Roman}$'s transit search. Our work serves as a quantitative baseline from which future dynamical studies can build to better understand the evolution of planetary systems in the Galactic Bulge.
\end{abstract}

\section{Introduction}
\label{sec:introduction}

The Nancy Grace Roman Space Telescope ($\textit{Roman}$) will soon begin its observational campaign \citep{Spergel2015, Akeson2019, schlieder2024survey}. As part of its Galactic Bulge Time-Domain Survey (GBTDS), $\textit{Roman}$ will extend the Galactic exoplanet census into the Milky Way's Bulge via microlensing and transit techniques \citep{2025arXiv250510574O}. These surveys are projected to discover an additional ${\sim}10^{4-5}$ transiting gas giants \citep[at orbital distances ${\lesssim}$au;][]{wilbarpow23}, and ${\sim}10^3$ wide-orbit \citep[][]{penny2019predictions} and/or free-floating \citep[][]{johnson2020predictions} microlensing planets \citep[with bound planet detectability maximized at projected separation ${\sim}1{-}5$ au;][]{gau12}. $\textit{Roman}$'s enlarged planet sample will revolutionize exoplanet demographics, and by extension, provide the first population-wide insights into the evolution of planetary systems in dense and kinematically hot stellar environments.

One major area of focus for \textit{Roman}'s microlensing survey is the study of ``free-floating planets'' (FFPs). Pre-\textit{Roman} estimates suggest that FFPs are remarkably common in the Galaxy, with as many as dozens per star \citep{sumi2023free}. The primary channel for their production, however, remains unclear: planet-planet scattering \citep{verray12, hadwu26}, stellar flybys \citep{yu2024free}, post-main sequence evolution \citep{veras2014great}, and stellar binaries \citep{col24, colder25} may each contribute to uncertain degrees. These mechanisms make different predictions for the proportion of planets that are truly gravitationally unbound from their star (i.e., ``rogue''), versus those merely stranded on long-period orbits (``detached'') outside of ${\gtrsim} 10\ \rm au$. Microlensing detections are unable to distinguish between these two cases, owing to the lack of host star signature in the microlensing event at separations ${\gtrsim}$10 au \citep{han2005microlensing, mroz2020free,yeeken25}. This uncertainty differentiating rogue vs. wide-orbit planets directly affects the interpretation of \textit{Roman\/}'s microlensing survey.

Another factor that will complicate the interpretation of \textit{Roman}'s survey is the extreme stellar environment of the Galactic Bulge. The typical stellar density (${\sim} 10 \ \rm stars \ pc^{-3}$) and velocity dispersion (${\sim} 100 \ \rm km \ s^{-1}$) in the region surveyed by the GBTDS are orders of magnitude higher than the solar neighborhood probed by $\textit{Kepler}$ \citep{bor16} and $\textit{TESS}$ \citep{ricwinvan15}. In contrast to planetary systems in the solar neighborhood, stars (and their planets) studied by the GBTDS therefore experience frequent, high-speed stellar flybys \citep[${\sim}10$ within 1000 au per Gyr;][see also \citealt{jimpiclak13}]{mctkipjoh20}, which may substantially perturb or eject planets at wide separations \citep[e.g.][]{brorei22}. In this Letter, we quantify to what extent such repeated, impulsive stellar flybys may contribute to the FFP population by truncating the outer limits of planetary systems. Our results therefore help clarify whether FFPs detected by \textit{Roman} are detached or rogue, by isolating the stellar flyby channel.

Our calculations also extend previous studies of stellar flybys to help fill a gap in parameter space unique to $\textit{Roman}$. Previous work has shown that stellar flybys may lead to a variety of dynamical outcomes, including producing planets on retrograde orbits \citep{breslau2019creating}, breaking resonant chains \citep{charalambous2025breaking}, ejecting cold Jupiters \citep{laughlin1998modification,spugieheg09}, and initiating high-eccentricity migration \citep[e.g.][]{hamtre17,rodet2021correlation, wang2022hot}. However, these studies do not address the relevant parameter space for the Galactic Bulge; namely, they assume low stellar densities, low velocity dispersions, or do not consider the effects of flybys accumulating over several Gyr. These less extreme conditions can amount to several orders of magnitude fewer close encounters of stellar perturbers than expected in the Galactic Bulge, or encounters that operate on qualitatively different timescales than in the Bulge. 

By isolating the effect of flybys on single-planet systems, our study provides a baseline calculation from which future studies may build to consider more complicated dynamical scenarios, such as multi-planet systems that subsequently undergo ejection \citep[e.g.][]{brorei22,hadwu26}, and perturbations from the Galactic tide \citep{veras2013exoplanets}. Our work also elucidates how flybys affect the orbital evolution of $\textit{bound}$ planets that are maximally detectable by $\textit{Roman}$'s microlensing survey (orbital separations ${\sim}1{-}10$ au). By extension, these results may impact interpretation of $\textit{Roman}$'s transiting planet sample, whose origins may stem from the evolution of planets further out. 
We demonstrate that bound planets superdiffuse through phase space, evolving qualitatively differently from planets outside the Bulge.

This Letter is structured as follows. In Section~\ref{sec:expectations}, we outline the basic quantitative expectations for flybys in the Galactic Bulge. In Section~\ref{sec:methods}, we describe our sampling procedure for Bulge stars and detail our $N$-body implementation. In Section~\ref{sec:results}, we analyze the outcomes of our flyby experiments in various parameter regimes. In Section~\ref{sec:discussion}, we provide an analytical model for planets undergoing repeated flybys and discuss the implications for free-floating planets and transit detections in the \textit{Roman} survey. We conclude in Section~\ref{sec:conclusion}.

\section{Basic Expectations for Stellar Flybys in the Galactic Bulge}\label{sec:expectations}

Before embarking on $N$-body calculations of planetary system stability in the Galactic Bulge, our goal here is to first establish the relevant timescales over which stellar flybys operate. We estimate the flyby frequency, distance of closest approach, and the timescale over which a given flyby occurs relative to planetary dynamical timescales. 

We assume Bulge stars exhibit a Maxwellian stellar velocity distribution with mass-independent one-dimensional dispersion $\sigma_{\star}$. With $n_{\star}$ the stellar volumetric number density, the average timescale between flybys inside a subject star's ``sphere of influence'' (also referred to as its ``encounter sphere'') of radius $d_{\rm enc}$ is \citep[averaging over the velocity distribution and neglecting gravitational focusing;][]{bintre08},

\begin{equation}\label{equation:t_flyby}
\begin{split}
    t_{\rm flyby}&\sim \bigg[4\sqrt{\pi}n_{\star}\sigma_{\star} d^{2}_{\rm enc}\bigg]^{-1},\\
    &\hspace{-0.8cm}\sim 10^{7}\bigg(\frac{n_{\star}}{19 \ \rm pc^{-3}}\bigg)^{-1}\bigg(\frac{\sigma_{\star}}{120 \ \rm km \ s^{-1}}\bigg)^{-1}\\
    &\times\bigg(\frac{d_{\rm enc}}{500 \ \rm au}\bigg)^{-2} \rm \ yr,
\end{split}
\end{equation}

\noindent where our choices for the stellar density and velocity dispersion are elaborated in Section \ref{subsec:parameters}. With these choices, Equation \ref{equation:t_flyby} indicates that over the ${\sim}10$ Gyr lifetime of a planetary system in the Bulge \citep[][]{haszasfeu20,sitnes20}, ${\sim}10^{3}$ stellar flybys occur at appreciable distances ${\sim}500$ au. We define the rate of flybys $\Gamma{=}1/t_{\rm flyby}$.

Over a planetary system's lifetime $t_{\rm age}$, the approximate distance of closest flyby is,

\begin{equation}\label{equation:d_min}
\begin{split}
    d_{\rm min}&\sim \bigg[4\sqrt{\pi}t_{\rm age}n_{\star}\sigma_{\star}\bigg]^{-1/2},\\&\sim 15 \bigg(\frac{t_{\rm age}}{10 \ \rm Gyr}\bigg)^{-1/2}\bigg(\frac{n_{\star}}{19 \ \rm pc^{-3}}\bigg)^{-1/2}\\
    &\times\bigg(\frac{\sigma_{\star}}{120 \ \rm km \ s^{-1}}\bigg)^{-1/2} \ \rm au. 
\end{split}
\end{equation}

\noindent Equation \ref{equation:d_min} indicates that, over the 10 Gyr lifetime of a Bulge star, it is plausible that a stellar flyby can pass close to planets that are within ${\sim}$10 au.\footnote{To be more quantitative, Poisson statistics dictate that the probability $\mathcal{P}$ that at least one encounter during the system's lifetime is within ${\leq}d_{\rm min}{=}20$ au is $\mathcal{P}{\sim}50\%$.}

Lastly, the timescale over which a flyby occurs (i.e. passes through the sphere of influence) is:

\begin{equation}\label{equation:t_passage}
\begin{split}
    t_{\rm passage}&\sim\frac{d_{\rm enc}}{\sigma_{\star}},\\
    &\sim 20 \bigg(\frac{d_{\rm enc}}{500 \ \rm au}\bigg)\bigg(\frac{\sigma_{\star}}{120 \ \rm km \ s^{-1}}\bigg)^{-1} \rm \ yr,
\end{split}
\end{equation}

\noindent indicating that although stellar passages are frequent, they are fast (``impulsive'') compared to planetary orbital timescales $P_{
\rm orb}{\sim}10^{3}\big(a/100 \ \mathrm {au}\big)^{3/2}(M_{\star}/M_{\odot})^{-1/2} \ \rm yr$. 

\begin{deluxetable*}{CCCCCCc}\label{tab:fiducial_table}
\tablecaption{Parameters used in our numerical flyby experiments.} 
\label{tab:parameters}
\tablecolumns{5}
\tablewidth{0pt}
\tablehead{
\colhead{Parameter} &
\colhead{Definition} &
\colhead{Value} & 
\colhead{Units} & 
}
\startdata
$M_{\rm max}$ & \rm maximum \ perturber \ mass & \{1, \ 10\} & M_{\odot} \\
$M_{\star}$ & \rm \ host \ star \ mass & \{0.3, \ 1\} & M_{\odot} \\
$a_0$ & \rm \ planet \ initial \ semi${-}$major \ axis & $[$10,500$]$ & \rm au \\
$e_0$ & \rm \ planet \ initial \ eccentricity & 0 & \nodata \\
$M_{\rm pl}$ & \rm \ planet \ mass & 10 & $M_{\oplus}$ \\
$n_{\star}$ & \rm \ stellar \ number \ density & $\{$19,50$\}$ & \rm pc$^{-3}$ \\
$\sigma_{\star}$ & \rm \ stellar \ velocity \ dispersion \ (1D) & \{80,120\} & \rm km \ s$^{-1}$ \\
$\Delta t$ & \rm \ integration \ time & [1,10] & \rm Gyr
\\
\enddata
\end{deluxetable*}

In summary, we estimate that stellar encounters in the Bulge occur frequently ($\Gamma{\sim}1 \rm \ flyby/10^{7}$ yr), at close distances (${\lesssim}500 \ \rm au$), but take place over ${\sim}1{-}2$ orders of magnitude shorter timescales than planetary orbital times (for orbits ${\gtrsim}30$ au). The goal of our Letter is to quantify how such frequent, but impulsive, flybys shape the outer limits of planetary systems (where planetary orbits are most strongly perturbed by stellar flybys).

\section{Methods}
\label{sec:methods}

Our methodology is described as follows. In Section \ref{subsec:sampling} we first detail our sampling procedure to simulate the effect of repeated stellar flybys on planetary systems. Section \ref{subsec:parameters} outlines our choices of stellar density and velocity dispersion, based on observations and modeling of the Galactic Bulge. Our numerical integration details are finally provided in Section \ref{subsec:numerics}. A summary of our parameter choices and setup is provided in Table \ref{tab:fiducial_table}.

\subsection{Simulating Flybys in the Bulge}\label{subsec:sampling}

Our sampling procedure to follow the effect of stellar flybys over a Bulge system's lifetime closely follows \cite{hamtre17}. We assume that the background distribution of stars surrounding a subject star is locally homogeneous, and it exhibits a one-dimensional velocity dispersion $\sigma_{\star}$ that is independent of mass. We follow stellar encounters that may perturb the subject system by seeding flyby stars near the encounter sphere (see Section \ref{subsec:numerics} for details on our numerical implementation of flyby initial conditions). Under the assumptions above, the differential flux of stars entering the encounter sphere, $dF$ (stars/area/time), reads \citep[see Section 3.2.1 of][]{hamtre17}:

\begin{equation}\label{equation:dF}
\begin{split}
    dF&=\frac{n_{\star}}{(2\pi\sigma_{\rm rel}^{2})^{3/2}}f(M_{\rm pert})dM_{\rm pert}\exp\bigg({\frac{GM}{d_{\rm enc}\sigma^{2}_{\rm rel}}}\bigg)\\
    & \times H(v_{\rm z})
    v_{\rm z}
    H\bigg(v^{2}-\frac{2GM}{d_{\rm enc}}\bigg)\exp\bigg(\frac{-v^{2}}{2\sigma^{2}_{\rm rel}}\bigg)dv_{\rm x}dv_{\rm y}dv_{\rm z}.
\end{split}
\end{equation}

\noindent where $\sigma_{\rm rel}{=}\sqrt{2}\sigma_{\star}$ is the relative velocity dispersion \cite[][]{bintre08}, $f(M_{\rm pert})dM_{\rm pert}$ is the fraction of perturber stars with masses in the interval $dM_{\rm pert}$ (computed using the Kroupa IMF; Equation \ref{equation:fm} below), $d_{\rm enc}$ is the radius of the encounter sphere, $M{=}M_{\star}{+}M_{\rm pert}$ is the sum of host star mass (we explore $M_{\star}{\in}\{0.3,1\} \ M_{\odot}$) and perturber mass $M_{\rm pert}$ (described below), $H$ is the Heaviside step function, and $v{=}|\mathbf{v}|{=}\sqrt{v^{2}_{\rm x}{+}v^{2}_{\rm y}{+}v^{2}_{\rm z}}$ is the perturber's initial velocity, written in a local coordinate system with $\mathbf{\hat{z}}$ pointed towards the host star. Equation \ref{equation:dF} describes the number of stars per unit time per unit area entering the encounter sphere, with velocity in the interval $dv_{\rm x}dv_{\rm y}dv_{\rm z}$. The first Heaviside function selects only stars that enter the encounter sphere, while the second keeps only stars with velocities above the system's escape speed.

Equation \ref{equation:dF} provides the cumulative distribution in hyperbolic excess velocity and periapse distance for perturbing stars. Neglecting gravitational focusing terms owing to the large stellar velocity dispersion we use, these cumulative distribution functions read:

\begin{equation}\label{equation:CDF}
\begin{split}
    \mathrm{CDF}(<d_{\rm peri})&\propto d^{2}_{\rm peri},\\
    \mathrm{CDF}(<v_{\infty})&=1-\exp\bigg(\frac{-v^{2}_{\infty}}{2\sigma^{2}_{\rm rel}}\bigg)\bigg[1+\frac{v^{2}_{\infty}}{2\sigma^{2}_{\rm rel}}\bigg].
\end{split}
\end{equation}

\noindent For each flyby, we sample the hyperbolic excess velocity $v_{\infty}$ and periapse distance $d_{\rm peri}$ from Equations \ref{equation:CDF}.\footnote{Equation \ref{equation:CDF} yields a different $v_{\infty}$ distribution than the pure Maxwellian used in other stellar flyby experiments \citep[e.g.][]{zinbatada20,brorei22}. The departure from a pure Maxwellian distribution in $v_{\infty}$ stems from Equation \ref{equation:dF}'s correction for the incoming flux of perturbers at the encounter sphere \citep{hen72}.}

\begin{figure*}
    \centering
    \includegraphics[width=1\linewidth]{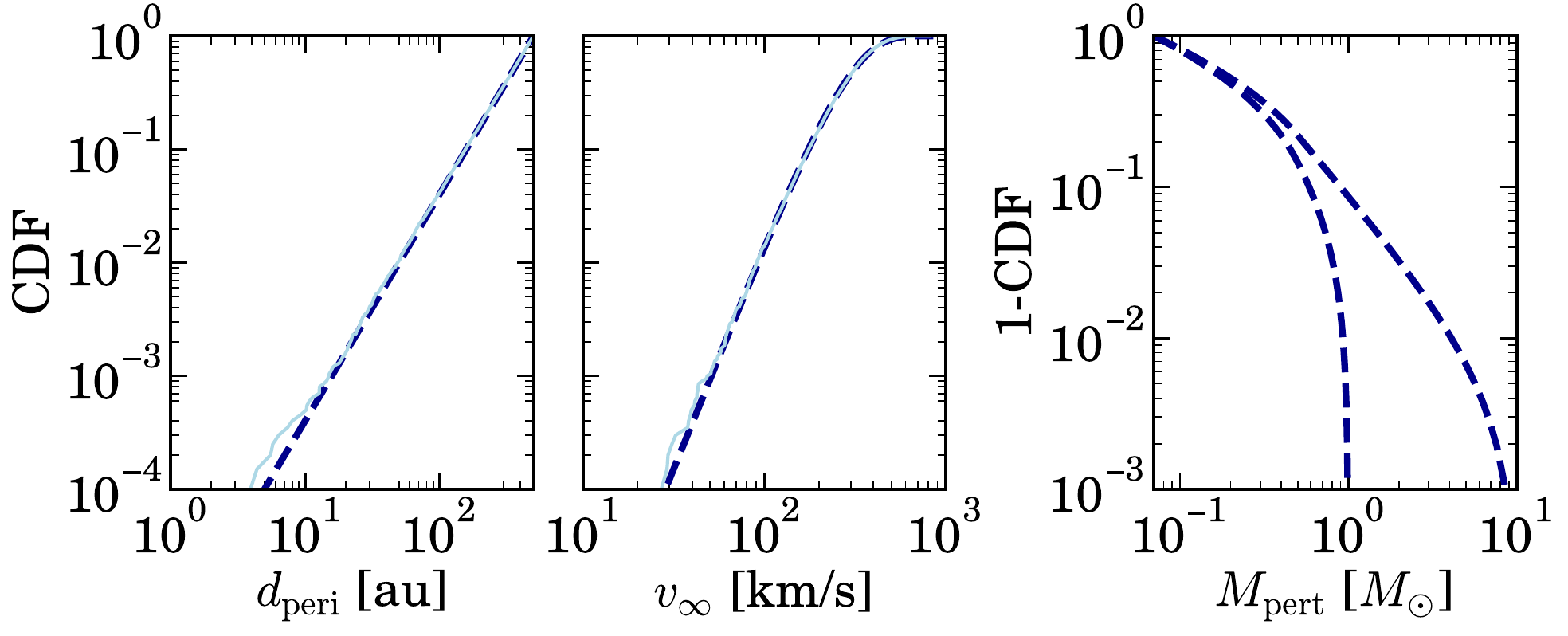}
    \caption{Cumulative distribution of flyby properties in our model of the Galactic Bulge (computed using $n_{\star}{=}19 \ \rm pc^{-3}$, $\sigma_{\star}{=}120 \ \rm km \ s^{-1}$, $d_{\rm enc}{=}500 \ \rm au$). Left and middle panels display the cumulative distributions in flyby periapse distances and hyperbolic excess velocities, respectively. Dark blue dashed curves show the analytic CDF from our stellar distribution \citep[Equation \ref{equation:dF}; adopted from][]{hamtre17}, while light blue solid show the Monte Carlo sample from our modified version of \texttt{AIRBALL}. The right panel shows the complementary cumulative distribution in perturber mass (from our Kroupa IMF), for our two choices of maximum perturber mass of 1 and 10 $M_{\odot}$ (left and rightmost curves). Stellar flybys in the Bulge are impulsive (large $v_{\infty}$) and distant (majority ${\gtrsim} 100 \ \rm au$), but also frequent ($\Gamma{\sim}1 \ \rm flyby/10^{7} \ \rm yr$).
    \label{fig:encounters}}
\end{figure*}

Perturber masses are sampled from the Kroupa IMF \citep[e.g.][]{kroweipfl13,krogjejer26}:

\begin{equation}\label{equation:fm}
\begin{split}
    f(M_{\rm pert})&\propto M^{-1.3}_{\rm pert}, \ \ 0.07 \leq M_{\rm pert} < 0.5 \ M_{\odot}\\
    & \propto M^{-2.3}_{\rm pert}, \ \ 0.5 \leq M_{\rm pert} < M_{\rm max},
\end{split}
\end{equation}

\noindent where $M_{\rm max}$ is an upper mass limit. Using $M_{\rm max}{=}1 \ M_{\odot}$ gives a mean $\langle M_{\rm pert} \rangle{=}0.3 \ M_{\odot}$. In this work we explore $M_{\rm max}{\in}\{1,10\} \ M_{\odot}$ to isolate the effect of rare but massive perturbers on the flyby statistics. While stars more massive than ${\gtrsim}1 \ M_{\odot}$ are expected to exit the main sequence after ${\lesssim}\rm a \ few$ Gyr, stellar remnants in the form of neutron stars and black holes are expected to populate the upper end of the perturber mass function \citep[e.g.][]{roslamlu22}. In this Letter we do not adopt a model for the stellar remnant mass function, but instead crudely emulate the massive remnant population by extending the stellar IMF up to $10 \ M_{\odot}$. We plan to address how more realistic stellar remnant mass functions impact flybys in the Bulge in a follow-up study. 

\subsection{Stellar Dynamical Parameters in the Bulge}\label{subsec:parameters}

Our study relies on two stellar dynamical parameters in the Bulge: stellar number density $n_{\star}$, and velocity dispersion $\sigma_{\star}$. We describe our choices for these parameters in turn below.

We estimate the stellar number density in $\textit{Roman}$'s field of view during the GBTDS using the $\texttt{GENULENS}$ tool \citep[][]{kosran22}.\footnote{\href{https://github.com/nkoshimoto/genulens}{Available here.}} $\texttt{GENULENS}$ uses the Galactic model of \cite{kosbabben21} to provide the statistical distribution of stars toward the Milky Way's Bulge. We leverage $\texttt{GENULENS}$ to simulate microlensing events in the GBTDS using a field of view at Galactic longitude/latitude $\{\ell,b\}{=}\{0.5,-1.4\} \ \deg$, respectively \citep[][]{2025arXiv250510574O}.\footnote{See also the $\textit{Roman}$ footprint provided in the user \href{https://shorturl.at/zkSFi}{ documentation.}} From the Galactic model we find an estimated stellar number density of $n_{\star}{=}19 \ \rm pc^{-3}$ at a distance of 8 kpc from the Sun. $\texttt{GENULENS}$ indeed predicts that ${\sim}70\%$ of stellar lenses belong to the Bulge at distance ${\sim}8$ kpc \citep[N. Koshimoto, private communication; see also][]{kosbensuz21}. 

To explore how variations in $n_{\star}$ affect our results, we repeat this exercise in $\textit{Roman}$'s second field of view toward Galactic center at $\{\ell,b\}{=}\{0,-0.5\}\rm \ deg$. This yields an estimate $n_{\star}{\sim}50 \ \rm pc^{-3}$. While $\textit{Roman}$'s field of view toward Galactic center has a higher stellar density, it is possible that microlensing detections there may be dominated by the nuclear stellar disk within the Bulge, which boasts $n_{\star}{\sim}500 \ \rm pc^{-3}$. We defer a study of flybys in the nuclear stellar disk to future work. We will find that the variations in $n_{\star}$ across the Bulge alone do not qualitatively change our conclusions.

We adopt two bracketing values of the stellar velocity dispersion in the Bulge of $\sigma_{\star}{=}\{80,120\}$ km s$^{-1}$, respectively (``slow'' and ``fast''). The upper limit of 120 km s$^{-1}$ is motivated by measurements across four regions of the Bulge \citep[$\{\ell,b\}{=}\{0,2\},\{0,-2\},\{1,-1\},\{-1,2\}\ \rm deg$;][]{valzocmuc18}, the Monte Carlo dynamical calculations of \cite{mctkipjoh20} for the innermost region of the Bulge (see their Figure 2), and the velocity dispersion of stars in Baade's window \citep[$\{\ell,b\}{=}\{1.13,-3.76\} \ \rm deg$;][]{bintre08}. The lower value of $\sigma_{\star}$ reflects estimates from the Galactic model of \cite{kosbabben21} (see their Equation 19 and Table 5), as well as \cite{mctkipjoh20} at a Galactocentric distance of ${<}$1 kpc.

\subsection{N-body Integrations $\&$ Sampling Procedure}\label{subsec:numerics}

Our flyby simulations use the open-source $N$-body code \texttt{REBOUND} \citep{rein2012rebound}. Since the periapse distance of a given flyby may be less than the semi-major axis of the planet, we use the adaptive timestep integrator \texttt{IAS15} to maintain numerical precision during such close encounters \citep{rein2015ias15}.

For each choice of maximum perturber mass $M_\text{max}$, host star mass $M_\star$, and stellar number density $n_\star$ and velocity dispersion $\sigma_{\star}$, we uniformly sample 50 initial semi-major axes $a_{0}{\in}[10,500]\ \rm au$, and 10 total integration times $\Delta t{\in} [1, 10 ]$ Gyr \footnote{We uniformly sample the integration times as a matter of implementation convenience; this choice does not affect our results.}. The total number of flybys $N_{\rm flyby}$ over a simulation time $\Delta t$ is assumed to be Poisson-distributed \citep[e.g.][]{baifab19}, drawn with mean $\lambda{=}\Gamma\Delta t$. We further assume zero initial eccentricity, and fix planet masses $M_{\rm pl}{=}10 \ M_{\oplus}$ (our results are not sensitive to $M_{\rm pl}$; see Appendix \ref{sec:mass-indep}). For a given set of integrations using $a_{0}$, we set $d_{\rm enc}{=}5 \ a_{0}$ to balance the need to resolve a large number of flybys for robust statistics ($N_{\rm flyby}{\propto}d^{2}_{\rm enc}$; Equation \ref{equation:t_flyby}) without wasting computational resources following weak flybys that do not affect planets (see Appendix \ref{sec:denc}). We repeat this sampling procedure 100 times for each choice of parameters, amounting to $50000$ integrations per stellar environment.  We record the statistics of our planet population (fraction ejected, and phase space distribution) as a function of time/number of flybys. 

\begin{figure*}
    \centering
    \includegraphics[width=1\linewidth]{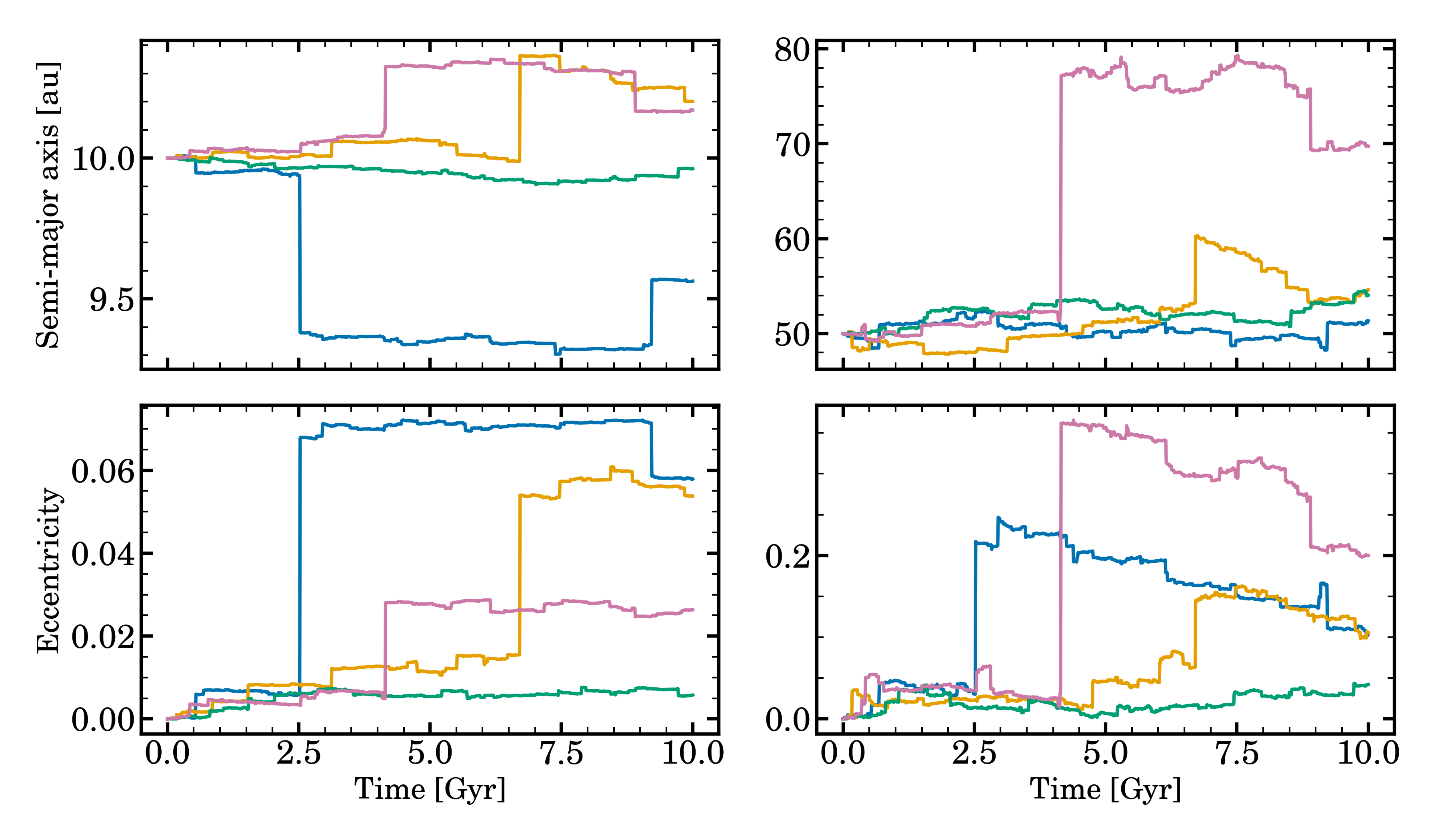}
    \caption{Evolution of semi-major axis (top row) and eccentricities (bottom row) of planets subject to repeated stellar flybys. Planets in the left column have initial semi-major axis $a_0{=}10\ \rm au$, and those in the right column have $a_0{=}50 \ \rm au$. The planets orbit M star hosts ($M_\star{=}0.3 \ M_\odot$) in a stellar environment with density $n_{\star}{=}19\ \rm{pc}^{-3}$. The maximum perturber mass for the flybys is $M_{\text{max}}{=}1 \ M_\odot$. Planet orbits random walk under the influence of impulsive but frequent stellar flybys.}
    \label{fig:time-series}
\end{figure*}

To sample the orbital and stellar parameters of each flyby, we use the open-source \texttt{AIRBALL} package \citep[][]{broreimoh24}. We modify \texttt{AIRBALL} to sample hyperbolic excess velocities from Equations \ref{equation:CDF} \citep[\texttt{AIRBALL} already samples periapse distance from our CDF following][]{zinbatada20}, and perturber masses from Equation \ref{equation:fm}. For each flyby, \texttt{AIRBALL} samples the inclination, argument of periapsis, and longitude of ascending node of the orbit by assuming the angular distribution of flybys is isotropic. The eccentricity of the orbit is then computed with $v_{\infty}$ and $d_{\rm peri}$.

From this orbit, \texttt{AIRBALL} begins the N-body integration with the flyby star at a distance of $10^{5}$ au (\texttt{AIRBALL} solves for the initial mean anomaly given this specified initial distance). We enforce this large initial distance to ensure a smooth ramp up in planet/flyby star interaction, starting from negligible interaction at this large separation.

\texttt{AIRBALL} also handles repeated stellar flybys by sequentially adding and removing perturbing stars from the \texttt{REBOUND} simulation. After a perturbing star completes its hyperbolic trajectory and exits the encounter sphere ($|r_{\rm pert}|{=}d_{\rm enc}$), it is removed from the simulation. Immediately afterwards, a new star is sampled, which is then added to the simulation. Since the period of the planet's orbit ($P_{\rm orb}{\sim}10^{3}(a/100 \ \mathrm{au})^{3/2}(M_{\star}/M_{\odot})^{-1/2}\ \rm yr$) is much shorter than the time for sequential flybys, the planet's mean anomaly is essentially random between subsequent encounters. 

This modeling choice reduces computational cost by neglecting long periods of integration between stellar flybys. We stress that this choice is only valid for single-planet systems on Keplerian orbits, for which the orbital shape and orientation remain constant between flybys. For multi-planet systems, secular timescales may be commensurate to or exceed the typical time between flybys, requiring full numerical integration over periods where no flybys occur.

\section{Results}\label{sec:results}

\subsection{Flyby Distributions}

We begin by summarizing the statistics of stellar flybys in our model of the Bulge. Figure \ref{fig:encounters} confirms our expectations from Section \ref{sec:expectations} that, on average, flybys are impulsive ($v_{\infty}{\sim}200$ km/s), distant ($d_{\rm peri}{\sim}100$s of au), and involve low mass perturbers. Figure \ref{fig:encounters} underscores however that a non-negligible fraction of flybys, ${\sim}1\%$, involve a massive perturber (when we allow $M_{\rm max}{=}10 \ M_{\odot}$) and/or a close flyby at ${\sim}$10s of au. We will find that the statistics of planet orbits under our flybys are dominated by such percent-level events. We observe good agreement between the sampled distributions in $\texttt{AIRBALL}$ and the analytic expectations.

\subsection{Statistics of Planetary Orbits}\label{subsec:statistics}

We now analyze the dynamical outcomes of single-planet systems subject to repeated flybys in the Galactic Bulge.

\Cref{fig:time-series} shows the time-series evolution of several planets' semi-major axes and eccentricities as a result of repeated stellar flybys. Each planet's evolution consists of many small perturbations to the orbital energy and eccentricity, and $1{-}2$ extreme perturbations; we discuss this behavior in more detail in Section~\ref{sec:levy}. We emphasize that even for $a_0{=}10\ \text{au}$, planets may attain non-trivial eccentricities and exchange significant orbital energy with the perturbing stars.

To survey how planetary orbital phase space evolves with flyby properties, we generate a series of contour maps depicting the ejection rate and the median eccentricity and inclination of the planets that remain bound for each ensemble. To maximize visual clarity, our contour maps apply a Gaussian blur to average each quantity with its nearest neighbors. 

\begin{figure*}
    \centering
    \includegraphics[width=\linewidth]{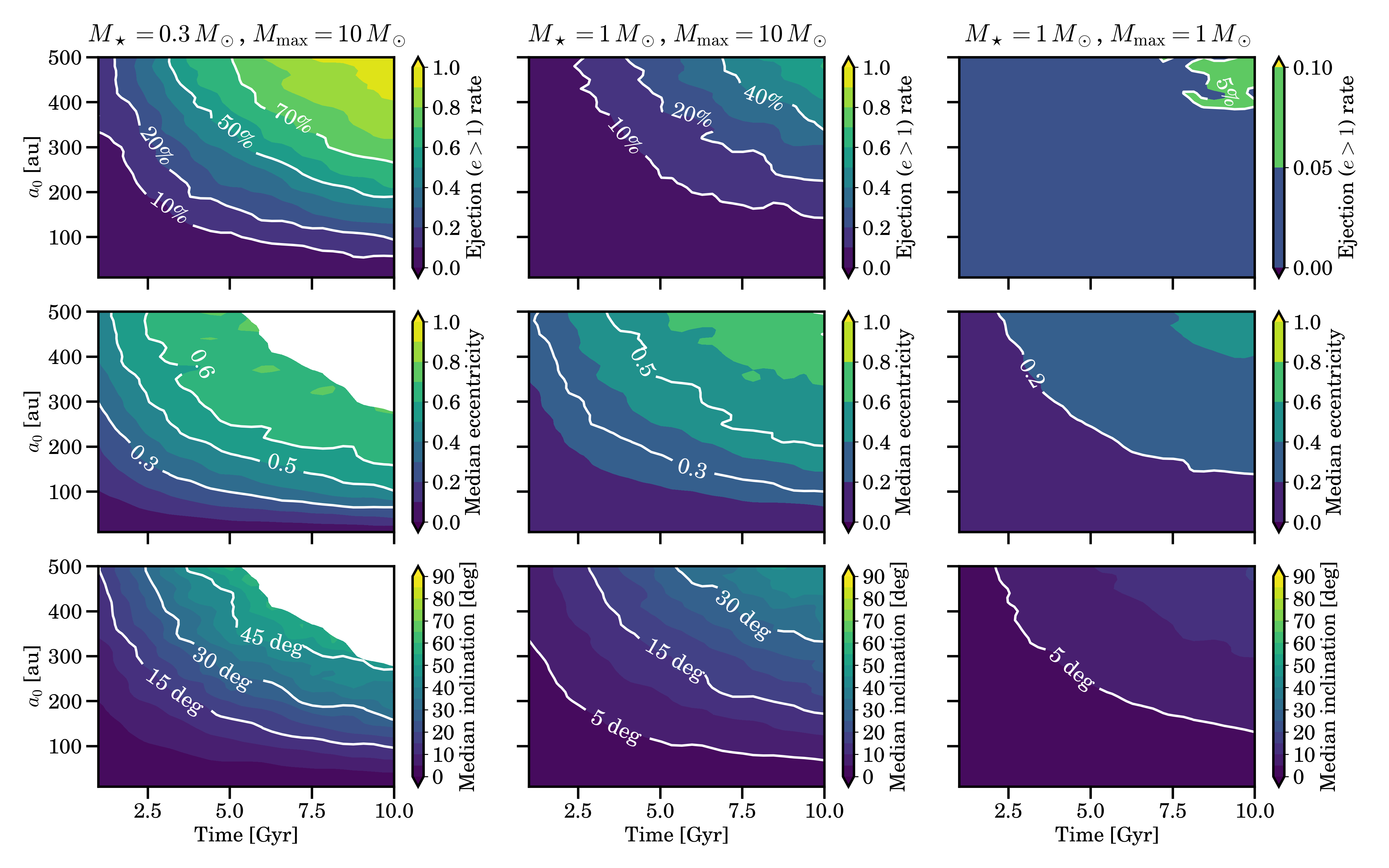}
    \caption{Distribution of planet ejection rate (top row), median eccentricity (middle row), and median inclination (bottom row) as a function of system age and initial semi-major axis $a_{0}$. From left to right, each column uses host star mass $M_{\star}$ and maximum perturber mass $M_{\max}$ of $\{M_{\star},M_{\rm max}\}{=}\{0.3,10\},\{1,10\},\{1,1\} \ M_{\odot}$, respectively. The stellar number density and velocity dispersion are $n_\star{=}19 \  \rm pc^{-3}$, and $\sigma_{\star}{=}120$ km s$^{-1}$, respectively. The majority of wide-orbit ($a_{0}{\gtrsim}250 \ \rm au$) planets orbiting M stars are ejected over ${\sim}$Gyr; the statistics of planetary orbits are dominated by the ${\sim}1\%$ of flybys involving a massive perturber.}
    \label{fig:low-density}
\end{figure*}

\begin{figure*}
    \centering
    \includegraphics[width=\linewidth]{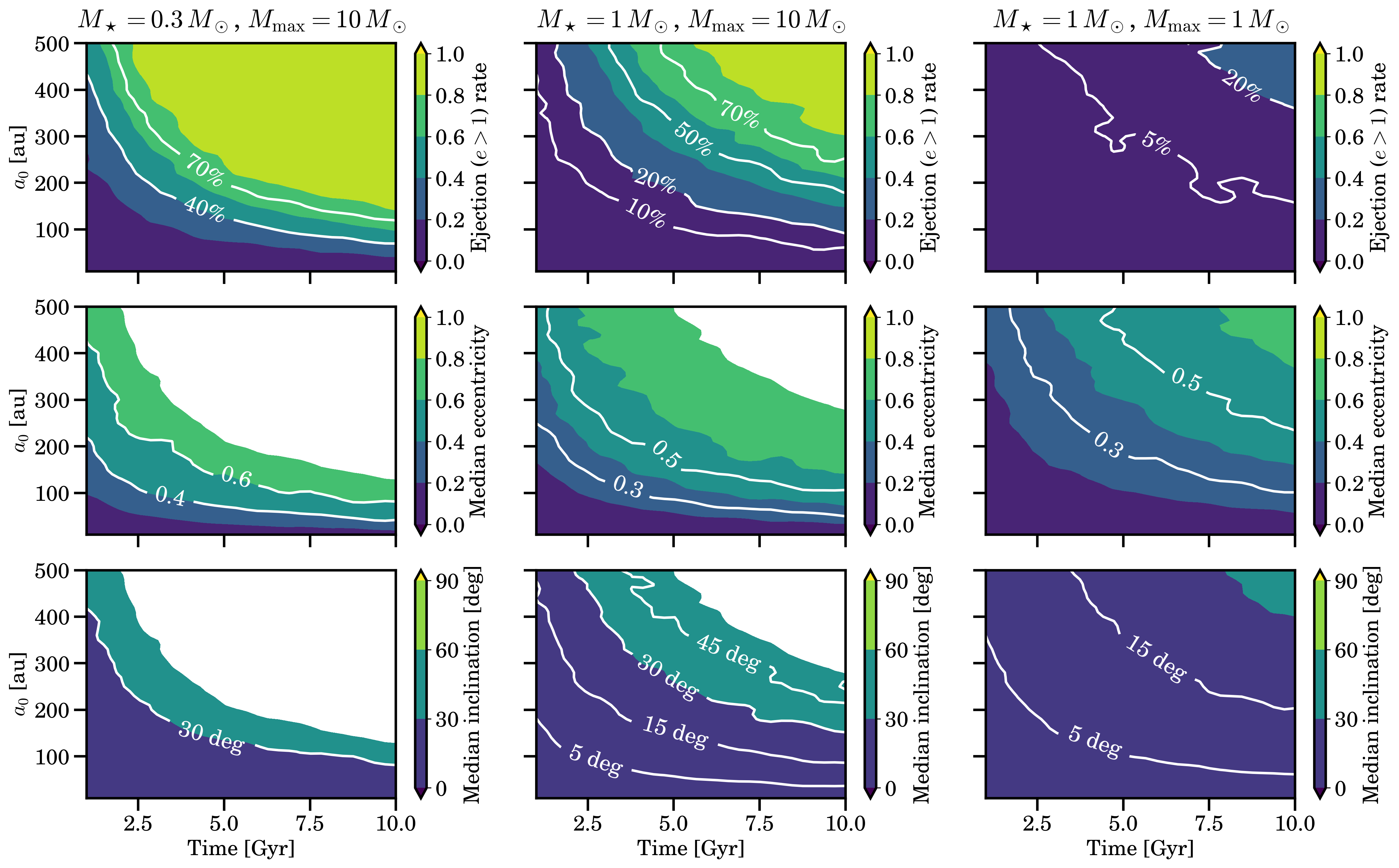}
    \caption{The same as \Cref{fig:low-density}, except here we use a higher stellar number density $n_\star{=}50\ \rm pc^{-3}$. Dynamical upheaval is enhanced due to the ${\sim}2.5{\times}$ larger number of flybys.}
    \label{fig:high-density}
\end{figure*}

\begin{figure*}
    \centering
    \includegraphics[width=\linewidth]{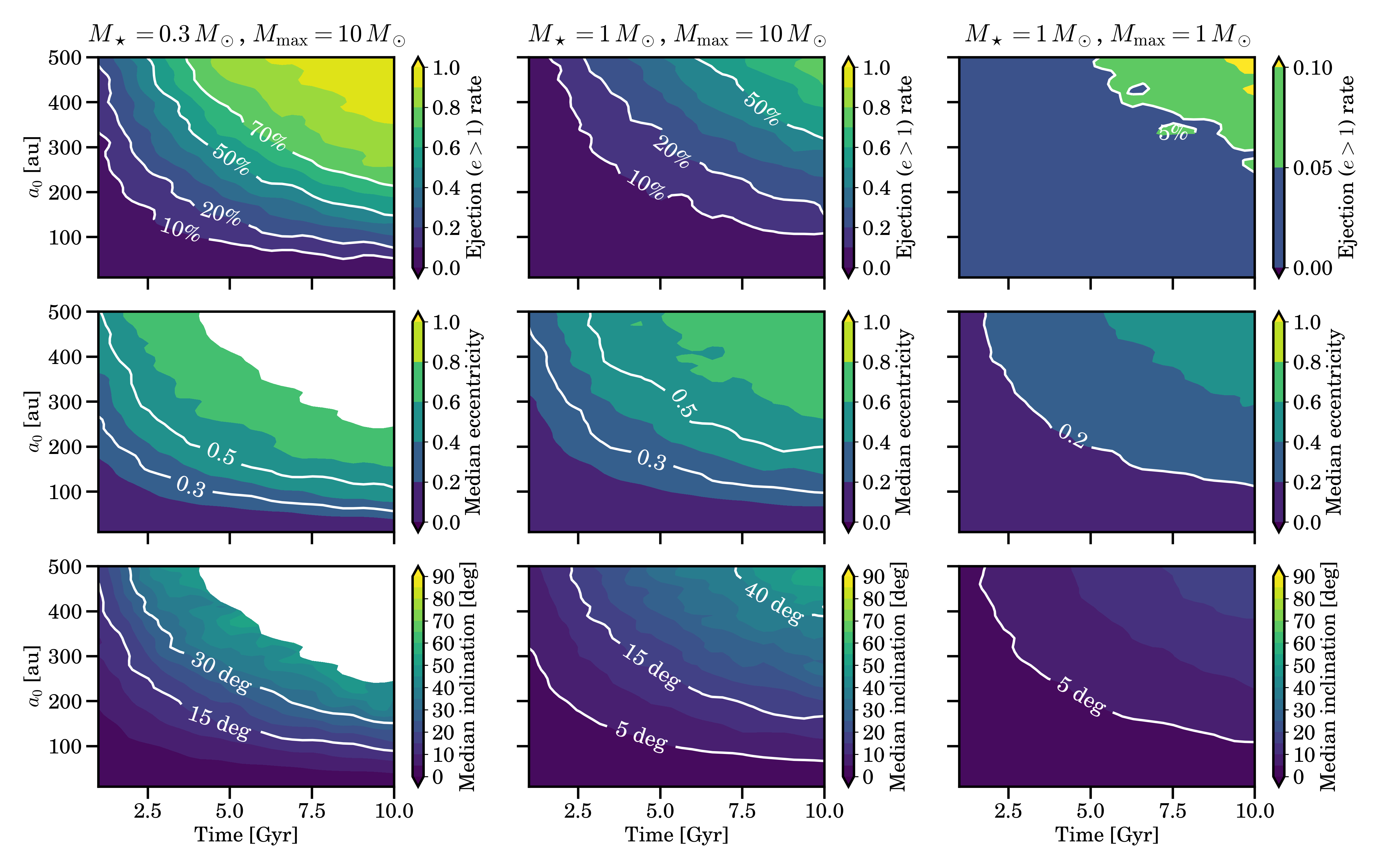}
    \caption{Exploring how the spread in stellar velocity dispersion affects the statistics of planetary orbits. This Figure is the same as \Cref{fig:low-density}, except we use a lower velocity dispersion $\sigma_{\star}{=}80\ \rm km \ s^{-1}$. Lower relative velocities produce fewer encounters that are less impulsive, ultimately leading to more dynamical upheaval than Figure \ref{fig:low-density}.}
    \label{fig:low-density-slow}
\end{figure*}

We first discuss planet ensembles with the fiducial stellar number density ($n_\star {=}19\ \rm pc^{-3}$) and the maximum perturber mass ($M_{\max}{=}10 \ M_{\odot}$). The distribution of ejection rates, median final eccentricities, and median inclinations as a function of $a_0$ and $\Delta t$ are shown for M star and solar mass hosts in the first two columns of \Cref{fig:low-density}. \Cref{fig:low-density} reveals that, in general, dynamical heating of planets takes billions of years; the fraction that attain high eccentricity grows monotonically over the 10 Gyr integration. We find that wide-orbit ($a_{0}{\gtrsim}250 \rm \ au$) planets orbiting M stars are affected relatively quickly, with the majority ejected over ${\sim}$a few Gyr. In contrast, only a minority of their close-in counterparts ($a_{0}{\sim} 50\ \text{au}$) are ejected over the entire 10 Gyr interval. Planets that avoid ejection do not survive unscathed however, but undergo substantial eccentricity and inclination pumping; 
planets grow to $e{\sim}0.7$, $i{\sim}45 \ \rm deg$ for far-out ($a_{0}{\gtrsim}250 \ \rm au$) initial orbits and up to $e{\sim}0.2$, $i{\sim}10 \ \rm deg$ for closer orbits ($a_{0}{\lesssim}100 \ \rm au$).
We conclude from \Cref{fig:low-density} that wide-orbit planets orbiting M stars experience substantial orbital upheaval. 

\Cref{fig:low-density} also highlights that the effect of flybys is lessened if host stars are solar mass (middle column), and can be muted almost entirely if the perturber mass function is further truncated to exclude massive stars (rightmost column). In cases with a 1 $M_{\odot}$ host star, ejection is still a common outcome, but generally only occurs for $a_0 {\gtrsim} 250\ \text{au}$ and requires ${\gtrsim}7$ Gyr. 
Limiting perturber masses to ${\leq}M_{\odot}$ on the other hand renders ejections rare, even for the widest orbits we consider; only such distant planets obtain mild eccentricities ($e{\sim} 0.2 $) and inclinations ($i {\sim} 5\ \text{deg}$). The upshot of this column in \Cref{fig:low-density} is that the ${\sim}1\%$ of flybys that involve a massive star (see \Cref{fig:encounters}) dominate the statistics of planet orbits. We further elucidate the strong dependence of dynamical outcomes on maximum perturber mass in Section \ref{sec:mmax-dependence}, and outline the physical reason behind the dominance of such percent-level events in Section \ref{sec:levy}.

\Cref{fig:high-density} summarizes how planetary orbital phase space differs in an environment with enhanced stellar number density ($n_\star{=}50\  \rm pc^{-3}$). Ejection rates, eccentricities and inclinations are substantially higher in this sample owing to the ${\sim}2.5{\times}$ larger number of flybys; unlike \Cref{fig:low-density}, ejection is the default outcome for wide-orbit planets around solar mass hosts (using heavy perturbers). Surviving planets retain substantial eccentricities/inclinations, even in the case with a truncated perturber mass. 

We next study the effect of a slower velocity dispersion on flyby outcomes, while retaining our fiducial stellar density. \Cref{fig:low-density-slow} showcases how our ``slow'' velocity dispersion (discussed in Section \ref{subsec:parameters}) alters the picture. The lesson of \Cref{fig:low-density-slow} is that, despite the fact that the number of flybys is decreased by ${\sim}2/3{\times}$ relative to our fiducial experiment, ejection rates and orbital elements in this sample are \textit{more} extreme. This modest enhancement in dynamical heating stems from an increased acceleration gradient experienced by the planet \citep{raymond2024future}.

The takeaway of Figures \ref{fig:high-density} and \ref{fig:low-density-slow} is that variations in stellar number density and velocity dispersion across the Galactic Bulge amount to order-unity corrections to the statistics of planetary orbits sculpted by flybys. The effect of these parameter choices is dwarfed by that of the perturber mass function, as dynamical heating is dominated by the ${\sim}1\%$ of stars more massive than 1 $M_{\odot}$.

\subsection{Dependence on Maximum Perturber Mass} \label{sec:mmax-dependence}

\begin{figure*}
    \centering
    \includegraphics[width=1\linewidth]{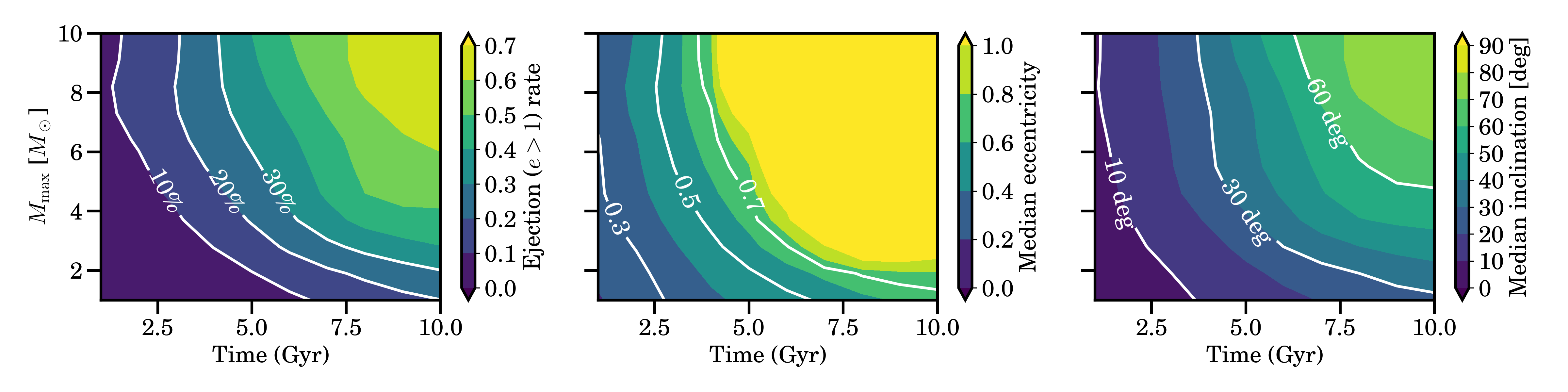}
    \caption{Ejection rate, median eccentricity, and median inclination for simulated systems as a function of integration time and maximum perturber mass. Planets are initialized with $a_0{=}250\ \text{au}$ around M star hosts ($M_\star{=}0.3 \ M_\odot$) in a stellar environment with density $n_\star{=}19\ \rm pc^{-3}$. Dynamical heating via flybys is enhanced as maximum perturber mass grows; perturbers more massive than ${\gtrsim}5 \ M_{\odot}$ produce similar levels of dynamical sculpting.}
    \label{fig:perturber-mass}
\end{figure*}

The results of Section \ref{subsec:statistics} indicate that the phase space explored by systems under flybys strongly depends on the maximum perturber mass. To quantify this dependence on perturber mass, \Cref{fig:perturber-mass} displays additional experiments varying the maximum perturber mass $M_{\max}{\in} [1,10] \ M_\odot$ (using $M_\star{=}0.3\ M_\odot$ and $a_0{=}250\ \text{au}$). Ejection rates and median post-flyby orbital elements grow strongly up to $M_{\max}{\sim}5 \ M_\odot$, above which the enhancement plateaus. Although the mass function of massive perturbers in the Bulge is poorly constrained \citep[e.g.][]{hegfry03,lulamdaw19}, we conclude from \Cref{fig:perturber-mass} that its \textit{functional \ form} (e.g. the power law slope) may be more important in setting the planetary orbital distribution than the upper mass limit, as perturbers ${\geq}5 \ M_{\odot}$ yield similar dynamical heating.

\section{Discussion}
\label{sec:discussion}

\subsection{Orbital Superdiffusion}

\label{sec:levy}
\begin{figure*}
    \centering
    \includegraphics[width=1\linewidth]{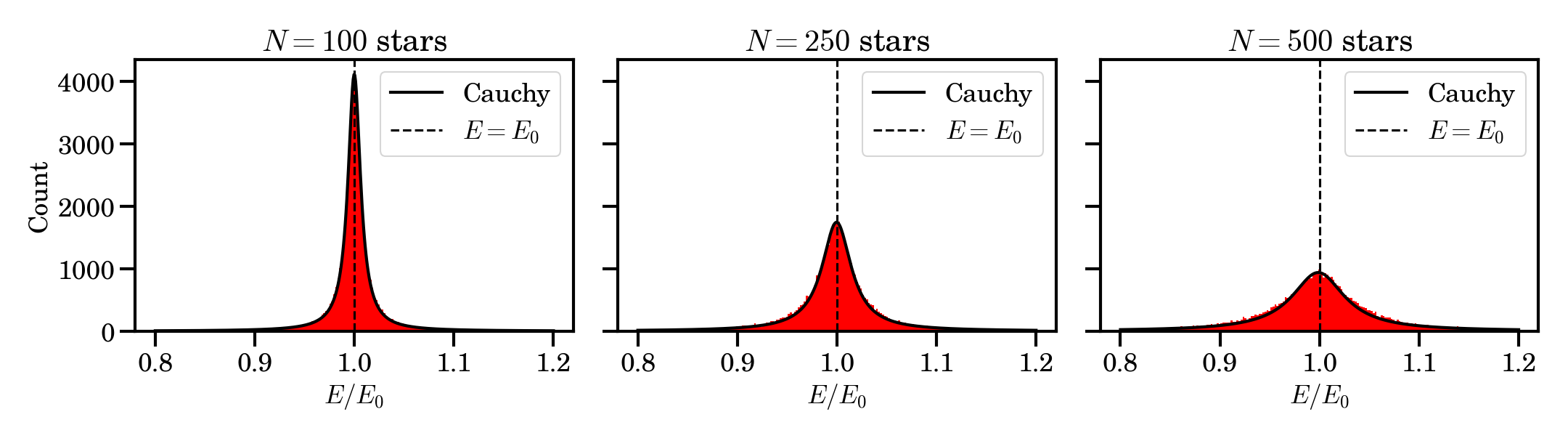}
    \caption{Distribution of normalized orbital energy $E(N)/E_0$ after $N{\in}\{100,250,500\}$ stellar flybys, from left to right panels. Vertical dashed lines correspond to $E{=}E_0$ (our initial condition). The distributions contain planets with initial semi-major axes $a_0{\in}[50,100]\ \rm au$, orbiting M stars ($M_\star{=}0.3 \ M_\odot$) in a stellar environment with $n_\star{=}19\ \rm pc^{-3}$. The maximum perturber mass is $M_{\text{max}}{=}1\ M_\odot$. Orbital energies are well-fit by a Cauchy (Lorentzian) distribution whose width grows quickly with time (black curves; see Figure \ref{fig:cauchy_width}), as expected if orbits evolve under a Lévy flight random walk. Planetary orbits in the Galactic Bulge ``superdiffuse'' under this random walk.}
    \label{fig:cauchy-fits}
\end{figure*}

\begin{figure}
\epsscale{1.15}
\plotone{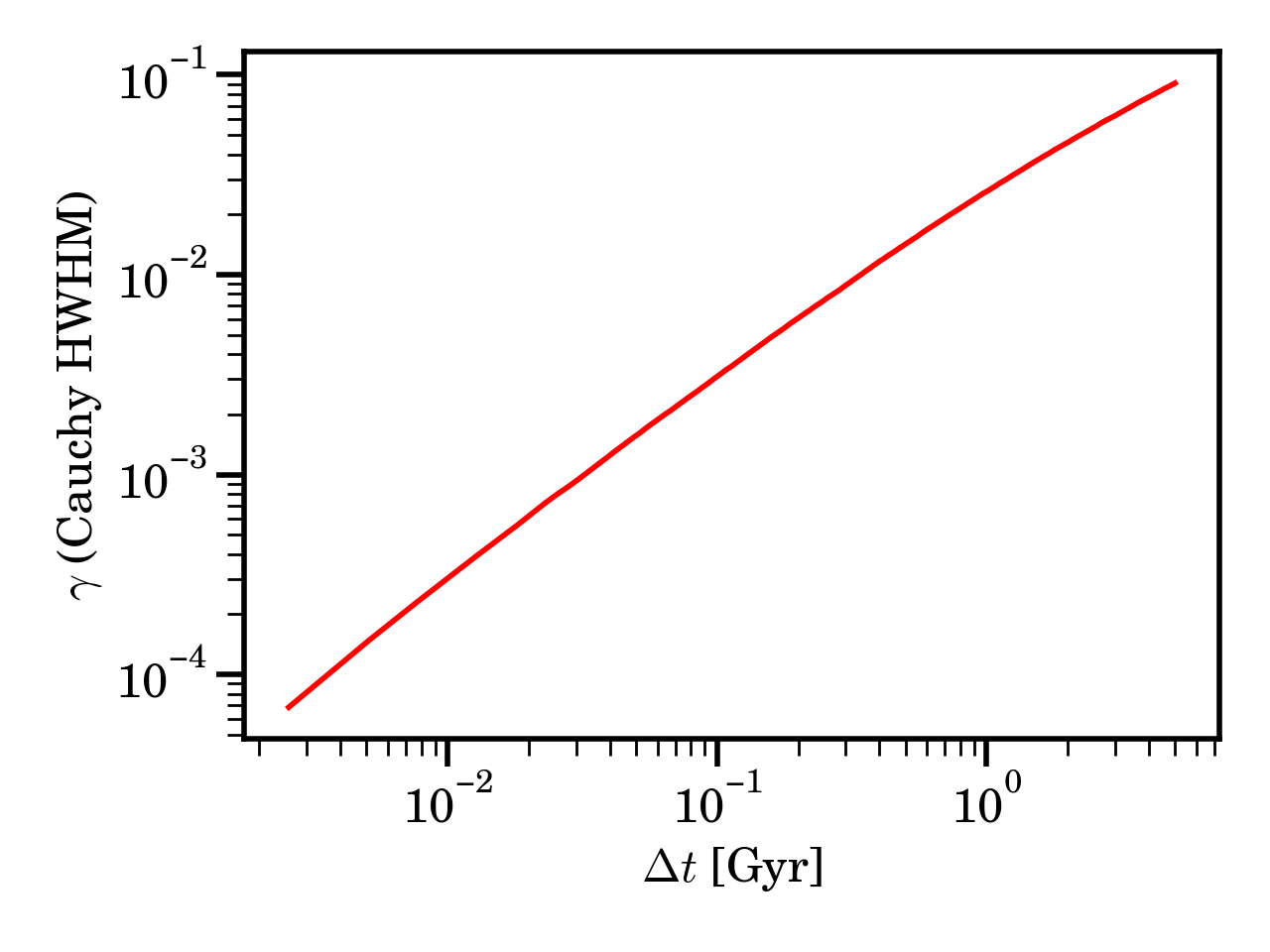}
\caption{The evolution in characteristic width $\gamma$ of our orbital energy distribution as a function of time (measured from Figure \ref{fig:cauchy-fits}). The distribution width grows approximately linearly with time, verifying the superdiffusive behavior of orbital energies. \label{fig:cauchy_width}}
\end{figure}

We have shown in Section \ref{subsec:statistics} and Figure \ref{fig:time-series} that planet orbits random walk under the influence of frequent, impulsive stellar flybys. We have also found that the statistics of planet orbits sculpted by flybys are dominated by the single strongest encounter, set by the maximum perturber mass (Figures \ref{fig:low-density} and  \ref{fig:perturber-mass}). Our goal here is to elucidate the physical driver of both this orbital diffusion, as well as the dominance of rare events. The physical intuition we develop for the evolution of planet orbits in the Bulge may help inform future, more sophisticated dynamical calculations. 

The orbital random walk showcased in Figure \ref{fig:time-series}, and the dominance of rare events shown in Figure \ref{fig:perturber-mass}, both stem from the fact that planets undergo a \textit{Lévy flight}. Lévy flights are a generalized random walk, roughly described as a sequence of minor perturbations punctuated by rare, large excursions \citep[e.g.][]{man83}. Lévy-type behavior arises across a variety of scales, from atmospheric eddy diffusion \citep{ric26}, particle transport in laminar, rotating fluids \citep{solweeswi93,shlzasfri95}, to cosmic ray propagation through the Galactic disk \citep{liaoh25}. Impulsive flybys similar to those explored in this paper are indeed known to produce Lévy flight behavior, as demonstrated analytically by \cite{colsar08} for the case of circular binary orbits \cite[see also][who suggest stellar flybys induce a Lévy flight but do not explicitly verify this]{brorei22}. Below we flesh out how our orbital evolution calculations can be boiled down to a simple model of Lévy flight dynamics.

Mathematically, Lévy flights arise when step sizes (or time increments) are drawn from a distribution with power law tails \citep[][]{zabdenkla15}, as opposed to standard, Gaussian step sizes \citep[e.g. in Brownian motion;][]{ein05}. In our study, these steps correspond to kicks in orbital energy $\epsilon$, which we denote by $\Delta\epsilon{=}\delta\epsilon/\epsilon$. As shown by \cite{colsar08}, heavy-tailed kicks arise during flybys because the frequency at which the planet energy (or angular momentum) changes by an amount $\Delta\epsilon$ is a power law. We visually confirm these heavy steps in Figure \ref{fig:time-series}, where orbital energy evolution is often dominated by a single extreme event. To illustrate the origin of our heavy-tailed step sizes, we determine how the frequency of flybys per change in relative energy, $d\Gamma/d\Delta\epsilon$, scales with energy change. Following \cite{spugieheg09}, the differential cross section $d\Sigma$ per change in relative energy is (for distant, impulsive flybys): 

\begin{equation}
    \frac{d\Sigma}{d\Delta \epsilon}=\frac{2\pi}{3}\frac{\sqrt{GM_{\rm pert}a^{3}}}{\sigma_{\star}(\Delta \epsilon)^{2}},
\end{equation}

\noindent such that $\frac{d\Sigma}{d\Delta\epsilon} d\Delta\epsilon$ is the cross section for achieving a relative energy change in $[\Delta\epsilon,\Delta\epsilon{+}d\Delta\epsilon]$ (and assuming $M_{\rm pert}{\sim}M_{\star}$). Armed with the cross section, we find that the frequency of flybys, $\Gamma{=}n_{\star}\sigma_{\star}\Sigma$, drops with the relative energy change like $d\Gamma/d\Delta\epsilon \propto 1/(\Delta\epsilon)^{2}$; the frequency of large energy excursions is small, but not vanishingly so.

Orbital diffusion under power-law step sizes differs from the standard random walk: the Central Limit Theorem no longer holds (the variance and higher order moments are infinite), and thus the sum of random steps does not tend toward a Gaussian. Lévy evolution is instead described by the Generalized Central Limit Theorem, i.e. the sum of steps drawn from a heavy-tailed distribution converges to a ``Lévy-stable distribution'' \citep[e.g.][]{gnekol68}. As a result of this difference in summing random variables, the variance in orbital energy displacement evolves like $\sqrt{\langle \Delta \epsilon^{2} \rangle}{\propto}t^{\xi/2}$, where $\xi$ is not in general equal to 1. The power law tail of step sizes sets this time evolution; in our case the Lévy flight yields \textit{superdiffusion}, with $\xi{=}2$.

Figure \ref{fig:cauchy-fits} illustrates this superdiffusion in orbital energy. To verify the diffusive behavior produced by our numerical calculations, we fit our normalized energy distribution, $f(E/E_{0})$ where $E_{0}$ is the initial energy, with a Cauchy (Lorentzian) function, as Cauchy distributions describe superdiffusive processes with linear growth in time \citep[][]{sat99}:

\begin{equation}\label{equation:cauchy}
    f(E/E_{0})=\frac{1}{\pi}\frac{\gamma}{(E/E_{0}-1)^{2}+\gamma^{2}},
\end{equation}

\noindent where $\gamma{=}\langle\Delta E/E_{0}\rangle$ sets the distribution width. We begin our planet ensemble as a delta function at $E/E_{0}{=}1$ and, as confirmed in Figure \ref{fig:cauchy-fits}, verify that a Cauchy distribution describes our data well as superdiffusion progresses.\footnote{In other words, the Cauchy distribution is the Green's function for the superdiffusion; the Fourier transform of the distribution in accumulated energy change is an exponentiated power law of index $\alpha{=}1$ \citep[e.g.][]{klazumshl95}.}

We diagnose this speed of superdiffusion by measuring the distribution width $\gamma$ as a function of time. As shown in Figure \ref{fig:cauchy_width}, we find an approximate scaling $\gamma{\propto}t$, in good agreement with the analytic expectation. We conclude that planet orbits undergo a bona fide Lévy flight, superdiffusing under the influence of stellar flybys.

We can use the superdiffusion model to estimate the characteristic diffusion/escape time for our planet ensembles. The orbital energy distribution $f$ evolves under superdiffusion via,

\begin{equation}
\partial_{t} f =\mathcal{D} \nabla^{\alpha}f,
\end{equation}

\noindent with $\nabla^{\alpha{=}1}$ the first-degree fractional Laplacian \citep[distinct from the standard 1D gradient, and sometimes denoted $-(-\Delta)^{\alpha/2}$ where $\Delta$ is the Laplacian;][]{poz16}, $x{=}E/E_{0}$, and $\mathcal{D}{=}\gamma(t)/t{=}\gamma_{0}/t_{0}$ the generalized diffusion coefficient (we have written $\gamma(t){=}\gamma_{0}(t/t_{0})$). From this equation the characteristic diffusion/escape time is (generalizing the standard $\tau{\sim}x^{2}/\mathcal{D}$ scaling):

\begin{equation}\label{equation:escape_time}
\begin{split}
\tau_{\rm esc} & \sim \frac{x_{\rm esc}}{\mathcal{D}},\\
& \sim \frac{x_{\rm esc}t_{0}}{\gamma_{0}},\\
& \sim 30 \bigg(\frac{\gamma_{0}/t_{0}}{3{\times}10^{-2}}\bigg)^{-1} \rm \ Gyr,
\end{split}
\end{equation}

\noindent where $x_{\rm esc}{\sim}1$, and in the last line we have evaluated $\gamma_{0}/t_{0}$ at $10^{-2}$ Gyr following Figure \ref{fig:cauchy_width}. Equation \ref{equation:escape_time} indicates that escape is negligible for the planet ensemble in Figures \ref{fig:cauchy-fits} and \ref{fig:cauchy_width} with $a_{0}{\in}[50,100]$ au, in good agreement with our numerical results (Figure \ref{fig:low-density}). Applying the same reasoning to a planet ensemble with $a_{0}{\sim}200$ au on the other hand yields an escape time $\tau_{\rm esc}{\sim}5$ Gyr, owing to the faster diffusion we measure with $\gamma_{0}/t_{0}{=}0.2$.

\subsection{Free-Floating Planets}

We have illustrated how stellar flybys in the Galactic Bulge routinely eject planets situated beyond ${\gtrsim}$hundreds of au, while planets at ${\sim}$tens of au remain bound. In this Section our goal is to explore how this distance-dependent ejection frequency may inform interpretation of ``free-floating'' planets amenable to discovery with $\textit{Roman}$. We outline how our results complement and extend theories for the origin of detached bodies probed by microlensing.

Even in the most extreme environments we consider (M star host, maximum number density $n_{\star}{=}50 \ \rm pc^{-3}$, and/or lower velocity dispersion $\sigma_{\star}{=}80 \ \rm km \ s^{-1}$), 
stellar flybys eject the majority of planets only when their initial semi-major axes are ${\gtrsim}$100 au.
In order for flybys to contribute to the free-floating planet population in the Bulge, planets must therefore be routinely emplaced at such wide separations. Populating such distant orbits during planet formation seems unlikely, as the radial extent of protoplanetary disks appears limited to ${\lesssim}$tens of au \citep[as measured in nearby star-forming regions; e.g.][]{zorbirros22,traroszha23}, with potentially even stronger disk truncation produced in the intense star formation conditions of the inner Milky Way \citep[][see also \citealt{hallee25}]{winkruche20}. 

Instead of primordial emplacement, wide-orbit ($\gtrsim$100 au) planets may arise via planet/planet scattering after formation \cite[e.g.][see also \cite{verray12} for further comparison between scattering theory and free-floating planet statistics]{chaformat08,jurtre08,vercrefor09}. \cite{hadwu26} in particular report that scattering between systems of 5 equal-mass planets typically strands several of the survivors far out, at ${\sim}10{-}100{\times}$ their initial separations (with ${\sim}1$ planet tightly bound, and another ejected). \cite{hadwu26} find that the degree to which the widest orbits are populated grows with the planets' masses and initial semi-major axes; for Neptune-like masses and an innermost planet initially located at $3 \ \rm au$, just ${\sim}10\%$ of their simulated planets lie at semi-major axes ${\gtrsim}100$ au by the end of evolution, whereas this fraction grows to ${\gtrsim}50\%$ if the initial location is 10 au (S. Hadden, private communication). Flybys may further increase the fraction of planets scattered outward by lifting their periastrons \citep[][]{baifab19}. Planet/planet scattering therefore seems the most promising channel by which planets in the Bulge may become susceptible to ejection via flybys.

Whether flybys can ultimately eject planets scattered to ${\sim}$hundreds of au depends on the timing of scattering and ejection: if planet/planet upheaval is too slow, flybys will not have a chance to subsequently induce ejection. We estimate whether these processes' timescales permit planet ejection via flybys in the following way. Stellar flybys require $\tau_{\rm esc}{\sim}5$ Gyr to evacuate the majority of wide-orbit planets. Meanwhile, \cite{hadwu26} find that the time for the mean number of bound planets to decrease by one is ${\sim}2.9{\times}10^{7}(m_{\rm p}/M_{\rm Neptune})^{-1.64}P_{1}$, where $P_{1}$ is the innermost planet's orbital period. We approximate this as the time for scattering to emplace planets at ${\gtrsim}$hundreds of au, $\tau_{\rm scatter}$. Requiring this scattering$\rightarrow$ejection scenario to be staged within a planetary system's lifetime, $\tau_{\rm life}{\sim}10$ Gyr, yields a limit on the innermost planet's initial orbital period:

\begin{equation}\label{equation:P1_scatter}
\begin{split}
    P_{1}&\leq\frac{\tau_{\rm life}-\tau_{\rm esc}}{2.9{\times}10^{7}}\bigg(\frac{m_{\rm p}}{m_{\rm Neptune}}\bigg)^{1.64},\\
    & \leq 172 \bigg(\frac{\delta \tau}{5 \ \rm Gyr}\bigg)\bigg(\frac{m_{\rm p}}{m_{\rm Neptune}}\bigg)^{1.64}\ \rm yr.
\end{split}
\end{equation}

\noindent with $\delta\tau{=}\tau_{\rm life}{-}\tau_{\rm esc}$. Equation \ref{equation:P1_scatter} tells us that, for a multiplanet system to scatter fast enough to then undergo ejection, the innermost planet must begin within 

\begin{equation}\label{equation:a1_scatter}
a_{1}{\lesssim}31\bigg(\frac{\delta \tau}{5 \ \mathrm{Gyr}}\bigg)^{2/3}\bigg(\frac{m_{\rm p}}{m_{\mathrm{Neptune}}}\bigg)^{1.09}\bigg(\frac{M_{\star}}{M_{\odot}}\bigg)^{1/3}\ \rm au.\end{equation}

\noindent On the other hand, \cite{hadwu26} also report that further-out Neptunes with $a_{1}{\gtrsim}$10 au are able to scatter more efficiently beyond ${\gtrsim}$100 au. Solving for the planet mass in Equation \ref{equation:a1_scatter} that yields $a_{1}{=}10$ au reveals that 
planets less massive than $m_{\rm p}{\lesssim}0.36 \ m_{\rm Neptune}{\sim}6 \ M_{\oplus}$ cannot reach the distant orbits susceptible to flyby ejection. We conclude that planets more massive than ${\gtrsim}6 \  M_{\oplus}$ may successfully scatter beyond ${\gtrsim}$100 au and undergo ejection via flybys over the system's lifetime. 

We note that scattering may be even more efficient at populating wide orbits than estimated here, due to the Lévy flight of planet orbits in the Bulge; it is conceivable that planets' superdiffusion through phase space enlarges the cross-section for instability. We elaborate on implications of the Lévy flight for inner planet orbits in Section \ref{subsec:transits}.

To summarize, our analysis indicates that planets emplaced beyond ${\gtrsim}$hundreds of au in the Bulge likely become unbound, rather than ``detached'' but bound as envisioned by \cite{hadwu26}. We stress however that this interpretation of ``free-floating'' planets as genuine rogues only applies to detections in the Bulge, where the stellar density is sufficient for flybys to evacuate the outer reaches of planetary systems.

If some FFPs are indeed the ejected products of outward planet/planet scattering, measurements of their mass distribution and occurrence rate may allow our model to constrain that of massive perturbers, since the ejection statistics are dominated by them. Backing out a clean constraint on the distribution of massive perturbers however may require a more secure understanding of what fraction of ejected planets stem from planet/planet scattering, versus other channels. Direct empirical constraints from $\textit{Roman}$ astrometric lensing measurements may also help clarify the upper end of the perturber mass function \citep[][]{lusinofe16,lulamdaw19,roslamlu22}.

\subsection{Implications for Transiting Planets}\label{subsec:transits}

Figure \ref{fig:time-series} illustrates that, although planets on relatively close orbits (${\lesssim}10$ au here) suffer no risk of ejection, they still superdiffuse through phase space. In this Section we outline how this constant orbital shuffling in the Galactic Bulge may qualitatively alter how planetary systems evolve, compared to those outside the Bulge. Specifically, we point out how orbital superdiffusion applies to interpretations of $\textit{Roman}$'s transit survey.

The $\textit{Roman}$ transit survey is projected to deliver ${\sim}10^{4-5}$ planets, largely dominated by hot Jupiters \citep[][]{wilbarpow23}. Superdiffusion may affect $\textit{Roman}$'s hot Jupiter yield in at least three ways: by directly torquing the orbits of hot Jupiter progenitors (Figure \ref{fig:time-series}), through the secular propagation of orbital disturbances inward to hot Jupiter progenitors \citep[e.g.][]{zaktre04}, or by affecting distant bodies thought to drive hot Jupiter formation via von Zeipel/Lidov/Kozai (ZLK) oscillations \citep[e.g.][]{lid62,koz62,wumur03}. These variations are unlikely to occur in isolation. For example, the random walk of a distant body's orbital inclination may quench or permit the operation of ZLK oscillations, which requires relative inclination to be maintained ${\sim}40{-}140 \ \deg$. In addition to a random walk in inclination, such perturbing bodies may also undergo ejection or eccentricity pumping under flybys \citep[as well as orbital perturbations from the Galactic tide; e.g.][]{stevigran24}; ejection would prohibit ZLK altogether, while a highly eccentric outer body may disrupt inner planetary systems. On the other hand, hot Jupiter formation via secular chaos \citep[e.g.][]{wulit11} may be enhanced when planet orbits superdiffuse in tandem with planet/planet interaction. The aggregate effect on the hot Jupiter population is not immediately obvious.

Our results therefore suggest that caution is warranted when applying standard orbital evolution theories to planets discovered by $\textit{Roman}$ in the Bulge. Theoretical interpretation of $\textit{Roman}$'s hot Jupiter sample will require future work to determine how the stability of inner planets is altered by superdiffusion. Another pressing avenue for future work is how ZLK migration is affected if the outer body undergoes a Lévy flight (e.g. are stellar binary companions long-term stable? Does their orbital random walk permit or prohibit ZLK oscillations?). We plan to address these issues in follow-up work.

The fact that planet orbits evolve qualitatively differently in the Bulge vs. solar neighborhood also suggests an opportunity to constrain planets' formation channels. To illustrate this idea, consider the case that $\textit{Roman}$ detects hot Jupiters with same frequency as those in the solar neighborhood. Determining how the various hot Jupiter formation channels differ between Bulge and solar neighborhood (e.g. whether ZLK is suppressed) may therefore constrain which must dominate to account for the same occurrence rate across both samples. 

\section{Conclusion}\label{sec:conclusion}

NASA's $\textit{Roman}$ Telescope is poised to expand the Galactic exoplanet census into the Milky Way's Bulge. In this work we explore how the dense stellar environment of the Bulge truncates the outer limits of planetary systems and shuffles planet orbits further in, via stellar flybys. We reiterate our conclusions below.

\begin{itemize}
\item Stellar flybys in the Bulge eject the majority of planets beyond ${\gtrsim}$250 au over ${\sim}$Gyr timescales. Planets inside ${\lesssim}$100 au undergo negligible ejection, even in the most extreme environments we consider. The fraction of planets ejected grows toward lower host star masses.
\item Planets' orbital evolution is dominated by the single strongest flyby over their lifetime; in our setup, this means that the fraction of planets ejected is dominated by the maximum perturbing star mass. Ejecting most planets beyond ${\gtrsim}$250 au requires a perturber mass function that extends above ${\gtrsim}1 \ M_{\odot}$.
\item Planet orbits in the Bulge superdiffuse through phase space (a generalized random walk). Even planets at no risk of ejection exhibit orbital superdiffusion, and can garner significant eccentricity and inclination excitation.
\item Planet/planet scattering at closer-in distances (${\sim}10$ au) may strand planets at ${\gtrsim}$hundreds of au quickly enough for flybys to subsequently eject them. Such planets ejected by flybys should be ${\gtrsim}6 \ M_{\oplus}$. Planet/planet scattering may therefore permit stellar flybys to contribute the free-floating planet population in the Bulge.
\end{itemize}

This paper is a step toward unraveling how the Galactic environment helps dictate the evolution of planetary systems. Our results suggest that planetary dynamics plays out qualitatively differently in the Galactic Bulge than the rest of the Milky Way; our understanding will soon improve when theory confronts observations from $\textit{Roman}$.

\section{Author Contributions}

T.H. conceived this project and methodology, and wrote the majority of the manuscript. D.L. developed and ran the numerical simulations including $\texttt{AIRBALL}$, analyzed the simulation data, assisted in methodology development and assisted in manuscript preparation. S.C.M. offered assistance in manuscript preparation and general feedback throughout this project.

\section{Acknowledgments}
D.L. thanks Fred Adams for helpful conversations. We also thank Garett Brown for feedback on $\texttt{AIRBALL}$, Sam Hadden for sharing simulation data, and Naoki Koshimoto for $\texttt{GENULENS}$ support. T.H. acknowledges enlightening exchanges with Andrew Vanderburg on microlensing/$\textit{Roman}$/free-floating planets. The raw dataset of our flyby experiments is available upon request. D.L. is supported by the National Science Foundation Graduate Research Fellowship Program. S.M. acknowledges support from the National Science Foundation under grant No. 2306391 and from the Alfred Sloan Foundation. We gratefully acknowledge access to computational resources through the MIT Engaging cluster at the Massachusetts Green High Performance Computing Center (MGHPCC) facility. Google Gemini and Claude Code were used to assist in figure formatting and for conversations regarding flyby literature; all figures were verified by the authors for consistency with the data.

\clearpage 

\bibliographystyle{aasjournal}
\bibliography{hallatt}

@ARTICLE{liaoh25,
       author = {{Liang}, Naixin and {Oh}, S. Peng},
        title = "{L{\'e}vy flights and leaky boxes: anomalous diffusion of cosmic rays}",
      journal = {\mnras},
         year = 2025,
        month = nov,
       volume = {543},
       number = {3},
        pages = {1911-1934},
          doi = {10.1093/mnras/staf1474},
archivePrefix = {arXiv},
       eprint = {2503.10747},
 primaryClass = {astro-ph.HE},
       adsurl = {https://ui.adsabs.harvard.edu/abs/2025MNRAS.543.1911L}
}

@BOOK{shlzasfri95,
       author = {{Shlesinger}, Micheal F. and {Zaslavsky}, George M. and {Frisch}, Uriel},
        title = "{L{\'e}vy Flights and Related Topics in Physics}",
         year = 1995,
       volume = {450},
       publisher= {Springer Berlin, Heidelberg},
       doi = {doi.org/10.1007/3-540-59222-9},
          doi = {10.1007/3-540-59222-9},
       adsurl = {https://ui.adsabs.harvard.edu/abs/1995LNP...450.....S}
}

@ARTICLE{ein05,
       author = {{Einstein}, A.},
        title = "{{\"U}ber die von der molekularkinetischen Theorie der W{\"a}rme geforderte Bewegung von in ruhenden Fl{\"u}ssigkeiten suspendierten Teilchen}",
      journal = {Annalen der Physik},
         year = 1905,
        month = jan,
       volume = {322},
       number = {8},
        pages = {549-560},
          doi = {10.1002/andp.19053220806},
       adsurl = {https://ui.adsabs.harvard.edu/abs/1905AnP...322..549E}
}

@ARTICLE{spugieheg09,
       author = {{Spurzem}, R. and {Giersz}, M. and {Heggie}, D.~C. and {Lin}, D.~N.~C.},
        title = "{Dynamics of Planetary Systems in Star Clusters}",
      journal = {\apj},
         year = 2009,
        month = may,
       volume = {697},
       number = {1},
        pages = {458-482},
          doi = {10.1088/0004-637X/697/1/458},
archivePrefix = {arXiv},
       eprint = {astro-ph/0612757},
 primaryClass = {astro-ph},
       adsurl = {https://ui.adsabs.harvard.edu/abs/2009ApJ...697..458S}
}

@INCOLLECTION{klazumshl95,
       author = {{Klafter}, J. and {Zumofen}, G. and {Shlesinger}, M.~F.},
        title = "{L{\'e}vy description of anomalous diffusion in dynamical systems}",
    booktitle = {L{\'e}vy Flights and Related Topics in Physics},
         year = 1995,
       editor = {{Shlesinger}, Micheal F. and {Zaslavsky}, George M. and {Frisch}, Uriel},
       publisher= {Springer Berlin, Heidelberg},
       volume = {450},
        pages = {196-215},
          doi = {10.1007/3-540-59222-9_35},
       adsurl = {https://ui.adsabs.harvard.edu/abs/1995LNP...450..196K}
}

@ARTICLE{colsar08,
       author = {{Collins}, Benjamin F. and {Sari}, Re'em},
        title = "{L{\'e}vy Flights of Binary Orbits due to Impulsive Encounters}",
      journal = {\aj},
         year = 2008,
        month = dec,
       volume = {136},
       number = {6},
        pages = {2552-2562},
          doi = {10.1088/0004-6256/136/6/2552},
archivePrefix = {arXiv},
       eprint = {0810.1525},
 primaryClass = {astro-ph},
       adsurl = {https://ui.adsabs.harvard.edu/abs/2008AJ....136.2552C}
}

@ARTICLE{zabdenkla15,
       author = {{Zaburdaev}, V. and {Denisov}, S. and {Klafter}, J.},
        title = "{L{\'e}vy walks}",
      journal = {Reviews of Modern Physics},
         year = 2015,
        month = apr,
       volume = {87},
       number = {2},
        pages = {483-530},
          doi = {10.1103/RevModPhys.87.483},
archivePrefix = {arXiv},
       eprint = {1410.5100},
 primaryClass = {cond-mat.stat-mech},
       adsurl = {https://ui.adsabs.harvard.edu/abs/2015RvMP...87..483Z}
}

@ARTICLE{jimpiclak13,
       author = {{Jim{\'e}nez-Torres}, Juan J. and {Pichardo}, B{\'a}rbara and {Lake}, George and {Segura}, Ant{\'\i}gona},
        title = "{Habitability in Different Milky Way Stellar Environments: A Stellar Interaction Dynamical Approach}",
      journal = {Astrobiology},
         year = 2013,
        month = may,
       volume = {13},
       number = {5},
        pages = {491-509},
          doi = {10.1089/ast.2012.0842},
archivePrefix = {arXiv},
       eprint = {1306.0464},
 primaryClass = {astro-ph.EP},
       adsurl = {https://ui.adsabs.harvard.edu/abs/2013AsBio..13..491J}
}

@ARTICLE{hadwu26,
       author = {{Hadden}, Sam and {Wu}, Yanqin},
        title = "{Free Floating or Merely Detached?}",
      journal = {\apj},
         year = 2026,
        month = may,
       volume = {1003},
       number = {1},
          eid = {70},
        pages = {70},
          doi = {10.3847/1538-4357/ae6508},
archivePrefix = {arXiv},
       eprint = {2507.08968},
 primaryClass = {astro-ph.EP},
       adsurl = {https://ui.adsabs.harvard.edu/abs/2026ApJ..1003...70H}
}

@ARTICLE{stevigran24,
       author = {{Stegmann}, Jakob and {Vigna-G{\'o}mez}, Alejandro and {Rantala}, Antti and {Wagg}, Tom and {Zwick}, Lorenz and {Renzo}, Mathieu and {van Son}, Lieke A.~C. and {de Mink}, Selma E. and {White}, Simon D.~M.},
        title = "{Close Encounters of Wide Binaries Induced by the Galactic Tide: Implications for Stellar Mergers and Gravitational-wave Sources}",
      journal = {\apjl},
         year = 2024,
        month = sep,
       volume = {972},
       number = {2},
          eid = {L19},
        pages = {L19},
          doi = {10.3847/2041-8213/ad70bb},
archivePrefix = {arXiv},
       eprint = {2405.02912},
 primaryClass = {astro-ph.GA},
       adsurl = {https://ui.adsabs.harvard.edu/abs/2024ApJ...972L..19S}
}

@ARTICLE{kosbensuz21,
       author = {{Koshimoto}, Naoki and {Bennett}, David P. and {Suzuki}, Daisuke and {Bond}, Ian A.},
        title = "{No Large Dependence of Planet Frequency on Galactocentric Distance}",
      journal = {\apjl},
         year = 2021,
        month = sep,
       volume = {918},
       number = {1},
          eid = {L8},
        pages = {L8},
          doi = {10.3847/2041-8213/ac17ec},
archivePrefix = {arXiv},
       eprint = {2108.11450},
 primaryClass = {astro-ph.EP},
       adsurl = {https://ui.adsabs.harvard.edu/abs/2021ApJ...918L...8K}
}

@ARTICLE{kosbabben21,
       author = {{Koshimoto}, Naoki and {Baba}, Junichi and {Bennett}, David P.},
        title = "{A Parametric Galactic Model toward the Galactic Bulge Based on Gaia and Microlensing Data}",
      journal = {\apj},
         year = 2021,
        month = aug,
       volume = {917},
       number = {2},
          eid = {78},
        pages = {78},
          doi = {10.3847/1538-4357/ac07a8},
archivePrefix = {arXiv},
       eprint = {2104.03306},
 primaryClass = {astro-ph.GA},
       adsurl = {https://ui.adsabs.harvard.edu/abs/2021ApJ...917...78K}
}

@software{kosran22,
       author = {{Koshimoto}, Naoki and {Ranc}, Cl{\'e}ment},
        title = "{nkoshimoto/genulens: Release version 1.2}",
         year = 2022,
        month = jul,
          eid = {10.5281/zenodo.6869520},
          doi = {10.5281/zenodo.6869520},
      version = {v1.2},
    publisher = {Zenodo},
       adsurl = {https://ui.adsabs.harvard.edu/abs/2022zndo...6869520K}
}

@ARTICLE{zaktre04,
       author = {{Zakamska}, Nadia L. and {Tremaine}, Scott},
        title = "{Excitation and Propagation of Eccentricity Disturbances in Planetary Systems}",
      journal = {\aj},
         year = 2004,
        month = aug,
       volume = {128},
       number = {2},
        pages = {869-877},
          doi = {10.1086/422023},
archivePrefix = {arXiv},
       eprint = {astro-ph/0404396},
 primaryClass = {astro-ph},
       adsurl = {https://ui.adsabs.harvard.edu/abs/2004AJ....128..869Z}
}

@ARTICLE{chaformat08,
       author = {{Chatterjee}, Sourav and {Ford}, Eric B. and {Matsumura}, Soko and {Rasio}, Frederic A.},
        title = "{Dynamical Outcomes of Planet-Planet Scattering}",
      journal = {\apj},
         year = 2008,
        month = oct,
       volume = {686},
       number = {1},
        pages = {580-602},
          doi = {10.1086/590227},
archivePrefix = {arXiv},
       eprint = {astro-ph/0703166},
 primaryClass = {astro-ph},
       adsurl = {https://ui.adsabs.harvard.edu/abs/2008ApJ...686..580C}
}

@ARTICLE{verray12,
       author = {{Veras}, Dimitri and {Raymond}, Sean N.},
        title = "{Planet-planet scattering alone cannot explain the free-floating planet population}",
      journal = {\mnras},
         year = 2012,
        month = mar,
       volume = {421},
       number = {1},
        pages = {L117-L121},
          doi = {10.1111/j.1745-3933.2012.01218.x},
archivePrefix = {arXiv},
       eprint = {1201.2175},
 primaryClass = {astro-ph.EP},
       adsurl = {https://ui.adsabs.harvard.edu/abs/2012MNRAS.421L.117V}
}

@ARTICLE{zorbirros22,
       author = {{Zormpas}, Apostolos and {Birnstiel}, Tilman and {Rosotti}, Giovanni P. and {Andrews}, Sean M.},
        title = "{A large population study of protoplanetary disks. Explaining the millimeter size-luminosity relation with or without substructure}",
      journal = {\aap},
         year = 2022,
        month = may,
       volume = {661},
          eid = {A66},
        pages = {A66},
          doi = {10.1051/0004-6361/202142046},
archivePrefix = {arXiv},
       eprint = {2202.01241},
 primaryClass = {astro-ph.EP},
       adsurl = {https://ui.adsabs.harvard.edu/abs/2022A&A...661A..66Z}
}

@ARTICLE{traroszha23,
       author = {{Trapman}, Leon and {Rosotti}, Giovanni and {Zhang}, Ke and {Tabone}, Beno{\^\i}t},
        title = "{How Large Is a Disk-What Do Protoplanetary Disk Gas Sizes Really Mean?}",
      journal = {\apj},
         year = 2023,
        month = sep,
       volume = {954},
       number = {1},
          eid = {41},
        pages = {41},
          doi = {10.3847/1538-4357/ace7d1},
archivePrefix = {arXiv},
       eprint = {2307.07600},
 primaryClass = {astro-ph.EP},
       adsurl = {https://ui.adsabs.harvard.edu/abs/2023ApJ...954...41T}
}

@BOOk{sat99,
       author = {{Sato}, K.},
        title = "{Levy Processes and Infinitely Divisible Distributions}",
      publisher = {Cambridge University Press},
         year = 1999,
}

@BOOK{poz16,
       author = {{Pozrikidis}, C.},
        title = "{The Fractional Laplacian}",
      publisher = {Boca Raton : CRC Press, Taylor $\&$ Francis Group},
         year = 2016,
}

@BOOK{gnekol68,
       author = {{Gnedenko}, B.~V. and {Kolmogorov}, A.~N.},
        title = "{Limit Distributions For Sums of Independent Random Variables}",
      publisher = {Reading, Mass., Addison-Wesley},
         year = 1968,
}

@BOOK{man83,
       author = {{Mandelbrot}, B.~B.},
        title = "{The fractal geometry of nature /Revised and enlarged edition/}",
         year = 1983,
         publisher = {San Francisco, W.H. Freeman},
       adsurl = {https://ui.adsabs.harvard.edu/abs/1983whf..book.....M}
}

@ARTICLE{solweeswi93,
       author = {{Solomon}, T.~H. and {Weeks}, Eric R. and {Swinney}, Harry L.},
        title = "{Observation of anomalous diffusion and L{\'e}vy flights in a two-dimensional rotating flow}",
      journal = {\prl},
         year = 1993,
        month = dec,
       volume = {71},
       number = {24},
        pages = {3975-3978},
          doi = {10.1103/PhysRevLett.71.3975},
       adsurl = {https://ui.adsabs.harvard.edu/abs/1993PhRvL..71.3975S}
}

@ARTICLE{ric26,
       author = {{Richardson}, Lewis F.},
        title = "{Atmospheric Diffusion Shown on a Distance-Neighbour Graph}",
      journal = {Proceedings of the Royal Society of London Series A},
         year = 1926,
        month = apr,
       volume = {110},
       number = {756},
        pages = {709-737},
          doi = {10.1098/rspa.1926.0043},
       adsurl = {https://ui.adsabs.harvard.edu/abs/1926RSPSA.110..709R}
}

@ARTICLE{brorei22,
       author = {{Brown}, Garett and {Rein}, Hanno},
        title = "{On the long-term stability of the Solar system in the presence of weak perturbations from stellar flybys}",
      journal = {\mnras},
         year = 2022,
        month = oct,
       volume = {515},
       number = {4},
        pages = {5942-5950},
          doi = {10.1093/mnras/stac1763},
archivePrefix = {arXiv},
       eprint = {2206.14240},
 primaryClass = {astro-ph.EP},
       adsurl = {https://ui.adsabs.harvard.edu/abs/2022MNRAS.515.5942B}
}

@ARTICLE{roslamlu22,
       author = {{Rose}, Sam and {Lam}, Casey Y. and {Lu}, Jessica R. and {Medford}, Michael and {Hosek}, Matthew W. and {Abrams}, Natasha S. and {Ramey}, Emily and {Vasylyev}, Sergiy S.},
        title = "{The Impact of Initial-Final Mass Relations on Black Hole Microlensing}",
      journal = {\apj},
         year = 2022,
        month = dec,
       volume = {941},
       number = {2},
          eid = {116},
        pages = {116},
          doi = {10.3847/1538-4357/aca09d},
archivePrefix = {arXiv},
       eprint = {2211.04471},
 primaryClass = {astro-ph.HE},
       adsurl = {https://ui.adsabs.harvard.edu/abs/2022ApJ...941..116R}
}

@ARTICLE{baifab19,
       author = {{Bailey}, Nora and {Fabrycky}, Daniel},
        title = "{Stellar Flybys Interrupting Planet-Planet Scattering Generates Oort Planets}",
      journal = {\aj},
         year = 2019,
        month = aug,
       volume = {158},
       number = {2},
          eid = {94},
        pages = {94},
          doi = {10.3847/1538-3881/ab2d2a},
archivePrefix = {arXiv},
       eprint = {1905.07044},
 primaryClass = {astro-ph.EP},
       adsurl = {https://ui.adsabs.harvard.edu/abs/2019AJ....158...94B}
}

@INPROCEEDINGS{krogjejer26,
       author = {{Kroupa}, Pavel and {Gjergo}, Eda and {Jerabkova}, Tereza and {Yan}, Zhiqiang},
        title = "{The initial mass function of stars}",
    booktitle = {Encyclopedia of Astrophysics, Volume 2},
         year = 2026,
       volume = {2},
        month = jan,
        pages = {173-210},
          doi = {10.1016/B978-0-443-21439-4.00035-3},
archivePrefix = {arXiv},
       eprint = {2410.07311},
 primaryClass = {astro-ph.GA},
       adsurl = {https://ui.adsabs.harvard.edu/abs/2026enap....2..173K}
}

@article{rein2012rebound,
  title={REBOUND: an open-source multi-purpose N-body code for collisional dynamics},
  author={Rein, Hanno and Liu, S-F},
  journal={Astronomy \& Astrophysics},
  volume={537},
  pages={A128},
  year={2012},
  publisher={EDP Sciences}
}

@article{rein2015ias15,
  title={IAS15: a fast, adaptive, high-order integrator for gravitational dynamics, accurate to machine precision over a billion orbits},
  author={Rein, Hanno and Spiegel, David S},
  journal={Monthly Notices of the Royal Astronomical Society},
  volume={446},
  number={2},
  pages={1424--1437},
  year={2015},
  publisher={Oxford University Press}
}

@misc{broreimoh24,
    title = {AIRBALL: a package for running and managing flybys using REBOUND},
    author = {{Brown}, Garett and {Rein}, Hanno and {Mohsin}, Hassan and {Chao-Ming Lam}, Ryan and {Generozov}, Aleksey and {He}, Linda and {Shi}, Ivy},
    year = {2024},
    url = {https://airball.gbrown.ca/},
}

@ARTICLE{zinbatada20,
       author = {{Zink}, Jon K. and {Batygin}, Konstantin and {Adams}, Fred C.},
        title = "{The Great Inequality and the Dynamical Disintegration of the Outer Solar System}",
      journal = {\aj},
         year = 2020,
        month = nov,
       volume = {160},
       number = {5},
          eid = {232},
        pages = {232},
          doi = {10.3847/1538-3881/abb8de},
archivePrefix = {arXiv},
       eprint = {2009.07296},
 primaryClass = {astro-ph.EP},
       adsurl = {https://ui.adsabs.harvard.edu/abs/2020AJ....160..232Z}
}

@ARTICLE{hamtre17,
       author = {{Hamers}, Adrian S. and {Tremaine}, Scott},
        title = "{Hot Jupiters Driven by High-eccentricity Migration in Globular Clusters}",
      journal = {\aj},
         year = 2017,
        month = dec,
       volume = {154},
       number = {6},
          eid = {272},
        pages = {272},
          doi = {10.3847/1538-3881/aa9926},
archivePrefix = {arXiv},
       eprint = {1710.00006},
 primaryClass = {astro-ph.EP},
       adsurl = {https://ui.adsabs.harvard.edu/abs/2017AJ....154..272H}
}

@ARTICLE{yeeken25,
       author = {{Yee}, Jennifer C. and {Kenyon}, Scott J.},
        title = "{Microlensing Constraints on the Stellar and Planetary Mass Functions}",
      journal = {\aj},
         year = 2025,
        month = aug,
       volume = {170},
       number = {2},
          eid = {132},
        pages = {132},
          doi = {10.3847/1538-3881/adeb84},
archivePrefix = {arXiv},
       eprint = {2503.11597},
 primaryClass = {astro-ph.EP},
       adsurl = {https://ui.adsabs.harvard.edu/abs/2025AJ....170..132Y}
}

@ARTICLE{gau12,
       author = {{Gaudi}, B. Scott},
        title = "{Microlensing Surveys for Exoplanets}",
      journal = {\araa},
         year = 2012,
        month = sep,
       volume = {50},
        pages = {411-453},
          doi = {10.1146/annurev-astro-081811-125518},
       adsurl = {https://ui.adsabs.harvard.edu/abs/2012ARA&A..50..411G}
}

@ARTICLE{Spergel2015,
       author = {{Spergel}, D. and {Gehrels}, N. and {Baltay}, C. and {Bennett}, D. and {Breckinridge}, J. and {Donahue}, M. and {Dressler}, A. and {Gaudi}, B.~S. and {Greene}, T. and {Guyon}, O. and {Hirata}, C. and {Kalirai}, J. and {Kasdin}, N.~J. and {Macintosh}, B. and {Moos}, W. and {Perlmutter}, S. and {Postman}, M. and {Rauscher}, B. and {Rhodes}, J. and {Wang}, Y. and {Weinberg}, D. and {Benford}, D. and {Hudson}, M. and {Jeong}, W.-S. and {Mellier}, Y. and {Traub}, W. and {Yamada}, T. and {Capak}, P. and {Colbert}, J. and {Masters}, D. and {Penny}, M. and {Savransky}, D. and {Stern}, D. and {Zimmerman}, N. and {Barry}, R. and {Bartusek}, L. and {Carpenter}, K. and {Cheng}, E. and {Content}, D. and {Dekens}, F. and {Demers}, R. and {Grady}, K. and {Jackson}, C. and {Kuan}, G. and {Kruk}, J. and {Melton}, M. and {Nemati}, B. and {Parvin}, B. and {Poberezhskiy}, I. and {Peddie}, C. and {Ruffa}, J. and {Wallace}, J.~K. and {Whipple}, A. and {Wollack}, E. and {Zhao}, F.},
        title = "{Wide-Field InfrarRed Survey Telescope-Astrophysics Focused Telescope Assets WFIRST-AFTA 2015 Report}",
      journal = {arXiv e-prints},
         year = 2015,
        month = mar,
          eid = {arXiv:1503.03757},
        pages = {arXiv:1503.03757},
          doi = {10.48550/arXiv.1503.03757},
archivePrefix = {arXiv},
       eprint = {1503.03757},
 primaryClass = {astro-ph.IM},
       adsurl = {https://ui.adsabs.harvard.edu/abs/2015arXiv150303757S}
}

@ARTICLE{Akeson2019,
       author = {{Akeson}, Rachel and {Armus}, Lee and {Bachelet}, Etienne and {Bailey}, Vanessa and {Bartusek}, Lisa and {Bellini}, Andrea and {Benford}, Dominic and {Bennett}, David and {Bhattacharya}, Aparna and {Bohlin}, Ralph and {Boyer}, Martha and {Bozza}, Valerio and {Bryden}, Geoffrey and {Calchi Novati}, Sebastiano and {Carpenter}, Kenneth and {Casertano}, Stefano and {Choi}, Ami and {Content}, David and {Dayal}, Pratika and {Dressler}, Alan and {Dor{\'e}}, Olivier and {Fall}, S. Michael and {Fan}, Xiaohui and {Fang}, Xiao and {Filippenko}, Alexei and {Finkelstein}, Steven and {Foley}, Ryan and {Furlanetto}, Steven and {Kalirai}, Jason and {Gaudi}, B. Scott and {Gilbert}, Karoline and {Girard}, Julien and {Grady}, Kevin and {Greene}, Jenny and {Guhathakurta}, Puragra and {Heinrich}, Chen and {Hemmati}, Shoubaneh and {Hendel}, David and {Henderson}, Calen and {Henning}, Thomas and {Hirata}, Christopher and {Ho}, Shirley and {Huff}, Eric and {Hutter}, Anne and {Jansen}, Rolf and {Jha}, Saurabh and {Johnson}, Samson and {Jones}, David and {Kasdin}, Jeremy and {Kelly}, Patrick and {Kirshner}, Robert and {Koekemoer}, Anton and {Kruk}, Jeffrey and {Lewis}, Nikole and {Macintosh}, Bruce and {Madau}, Piero and {Malhotra}, Sangeeta and {Mandel}, Kaisey and {Massara}, Elena and {Masters}, Daniel and {McEnery}, Julie and {McQuinn}, Kristen and {Melchior}, Peter and {Melton}, Mark and {Mennesson}, Bertrand and {Peeples}, Molly and {Penny}, Matthew and {Perlmutter}, Saul and {Pisani}, Alice and {Plazas}, Andr{\'e}s and {Poleski}, Radek and {Postman}, Marc and {Ranc}, Cl{\'e}ment and {Rauscher}, Bernard and {Rest}, Armin and {Roberge}, Aki and {Robertson}, Brant and {Rodney}, Steven and {Rhoads}, James and {Rhodes}, Jason and {Ryan}, Jr., Russell and {Sahu}, Kailash and {Sand}, David and {Scolnic}, Dan and {Seth}, Anil and {Shvartzvald}, Yossi and {Siellez}, Karelle and {Smith}, Arfon and {Spergel}, David and {Stassun}, Keivan and {Street}, Rachel and {Strolger}, Louis-Gregory and {Szalay}, Alexander and {Trauger}, John and {Troxel}, M.~A. and {Turnbull}, Margaret and {van der Marel}, Roeland and {von der Linden}, Anja and {Wang}, Yun and {Weinberg}, David and {Williams}, Benjamin and {Windhorst}, Rogier and {Wollack}, Edward and {Wu}, Hao-Yi and {Yee}, Jennifer and {Zimmerman}, Neil},
        title = "{The Wide Field Infrared Survey Telescope: 100 Hubbles for the 2020s}",
      journal = {arXiv e-prints},
         year = 2019,
        month = feb,
          eid = {arXiv:1902.05569},
        pages = {arXiv:1902.05569},
          doi = {10.48550/arXiv.1902.05569},
archivePrefix = {arXiv},
       eprint = {1902.05569},
 primaryClass = {astro-ph.IM},
       adsurl = {https://ui.adsabs.harvard.edu/abs/2019arXiv190205569A}
}

@ARTICLE{koz62,
       author = {{Kozai}, Yoshihide},
        title = "{Secular perturbations of asteroids with high inclination and eccentricity}",
      journal = {\aj},
         year = 1962,
        month = nov,
       volume = {67},
        pages = {591-598},
          doi = {10.1086/108790},
       adsurl = {https://ui.adsabs.harvard.edu/abs/1962AJ.....67..591K}
}

@ARTICLE{lid62,
       author = {{Lidov}, M.~L.},
        title = "{The evolution of orbits of artificial satellites of planets under the action of gravitational perturbations of external bodies}",
      journal = {\planss},
         year = 1962,
        month = oct,
       volume = {9},
       number = {10},
        pages = {719-759},
          doi = {10.1016/0032-0633(62)90129-0},
       adsurl = {https://ui.adsabs.harvard.edu/abs/1962P&SS....9..719L}
}

@ARTICLE{wumur03,
       author = {{Wu}, Y. and {Murray}, N.},
        title = "{Planet Migration and Binary Companions: The Case of HD 80606b}",
      journal = {\apj},
         year = 2003,
        month = may,
       volume = {589},
       number = {1},
        pages = {605-614},
          doi = {10.1086/374598},
archivePrefix = {arXiv},
       eprint = {astro-ph/0303010},
 primaryClass = {astro-ph},
       adsurl = {https://ui.adsabs.harvard.edu/abs/2003ApJ...589..605W}
}

@ARTICLE{jurtre08,
       author = {{Juri{\'c}}, Mario and {Tremaine}, Scott},
        title = "{Dynamical Origin of Extrasolar Planet Eccentricity Distribution}",
      journal = {\apj},
         year = 2008,
        month = oct,
       volume = {686},
       number = {1},
        pages = {603-620},
          doi = {10.1086/590047},
archivePrefix = {arXiv},
       eprint = {astro-ph/0703160},
 primaryClass = {astro-ph},
       adsurl = {https://ui.adsabs.harvard.edu/abs/2008ApJ...686..603J}
}

@ARTICLE{wulit11,
       author = {{Wu}, Yanqin and {Lithwick}, Yoram},
        title = "{Secular Chaos and the Production of Hot Jupiters}",
      journal = {\apj},
         year = 2011,
        month = jul,
       volume = {735},
       number = {2},
          eid = {109},
        pages = {109},
          doi = {10.1088/0004-637X/735/2/109},
archivePrefix = {arXiv},
       eprint = {1012.3475},
 primaryClass = {astro-ph.EP},
       adsurl = {https://ui.adsabs.harvard.edu/abs/2011ApJ...735..109W}
}

@BOOK{bintre08,
       author = {{Binney}, James and {Tremaine}, Scott},
        title = "{Galactic Dynamics: Second Edition}",
        publisher = {Princeton University Press},
         year = 2008,
       adsurl = {https://ui.adsabs.harvard.edu/abs/2008gady.book.....B}
}

@ARTICLE{winkruche20,
       author = {{Winter}, Andrew J. and {Kruijssen}, J.~M. Diederik and {Chevance}, M{\'e}lanie and {Keller}, Benjamin W. and {Longmore}, Steven N.},
        title = "{Prevalent externally driven protoplanetary disc dispersal as a function of the galactic environment}",
      journal = {\mnras},
         year = 2020,
        month = jan,
       volume = {491},
       number = {1},
        pages = {903-922},
          doi = {10.1093/mnras/stz2747},
       adsurl = {https://ui.adsabs.harvard.edu/abs/2020MNRAS.491..903W}
}

@INCOLLECTION{kroweipfl13,
       author = {{Kroupa}, Pavel and {Weidner}, Carsten and {Pflamm-Altenburg}, Jan and {Thies}, Ingo and {Dabringhausen}, J{\"o}rg and {Marks}, Michael and {Maschberger}, Thomas},
        title = "{The Stellar and Sub-Stellar Initial Mass Function of Simple and Composite Populations}",
    booktitle = {Planets, Stars and Stellar Systems. Volume 5: Galactic Structure and Stellar Populations},
         year = 2013,
        publisher = {Springer Science + Business Media Dordrecht},
       editor = {{Oswalt}, Terry D. and {Gilmore}, Gerard},
       volume = {5},
        pages = {115},
          doi = {10.1007/978-94-007-5612-0_4},
       adsurl = {https://ui.adsabs.harvard.edu/abs/2013pss5.book..115K}
}

@ARTICLE{wilbarpow23,
       author = {{Wilson}, Robert F. and {Barclay}, Thomas and {Powell}, Brian P. and {Schlieder}, Joshua and {Hedges}, Christina and {Montet}, Benjamin T. and {Quintana}, Elisa and {Mcdonald}, Iain and {Penny}, Matthew T. and {Espinoza}, N{\'e}stor and {Kerins}, Eamonn},
        title = "{Transiting Exoplanet Yields for the Roman Galactic Bulge Time Domain Survey Predicted from Pixel-level Simulations}",
      journal = {\apjs},
         year = 2023,
        month = nov,
       volume = {269},
       number = {1},
          eid = {5},
        pages = {5},
          doi = {10.3847/1538-4365/acf3df},
archivePrefix = {arXiv},
       eprint = {2305.16204},
 primaryClass = {astro-ph.EP},
       adsurl = {https://ui.adsabs.harvard.edu/abs/2023ApJS..269....5W}
}

@ARTICLE{bor16,
       author = {{Borucki}, William J.},
        title = "{KEPLER Mission: development and overview}",
      journal = {Reports on Progress in Physics},
         year = 2016,
        month = mar,
       volume = {79},
       number = {3},
          eid = {036901},
        pages = {036901},
          doi = {10.1088/0034-4885/79/3/036901},
       adsurl = {https://ui.adsabs.harvard.edu/abs/2016RPPh...79c6901B}
}

@ARTICLE{hallee25,
       author = {{Hallatt}, Tim and {Lee}, Eve J.},
        title = "{On the Formation of Planets in the Milky Way's Thick Disk}",
      journal = {\apj},
         year = 2025,
        month = feb,
       volume = {979},
       number = {2},
          eid = {120},
        pages = {120},
          doi = {10.3847/1538-4357/ad9aa1},
archivePrefix = {arXiv},
       eprint = {2408.09319},
 primaryClass = {astro-ph.EP},
       adsurl = {https://ui.adsabs.harvard.edu/abs/2025ApJ...979..120H}
}

@ARTICLE{ricwinvan15,
       author = {{Ricker}, George R. and {Winn}, Joshua N. and {Vanderspek}, Roland and {Latham}, David W. and {Bakos}, G{\'a}sp{\'a}r {\'A}. and {Bean}, Jacob L. and {Berta-Thompson}, Zachory K. and {Brown}, Timothy M. and {Buchhave}, Lars and {Butler}, Nathaniel R. and {Butler}, R. Paul and {Chaplin}, William J. and {Charbonneau}, David and {Christensen-Dalsgaard}, J{\o}rgen and {Clampin}, Mark and {Deming}, Drake and {Doty}, John and {De Lee}, Nathan and {Dressing}, Courtney and {Dunham}, Edward W. and {Endl}, Michael and {Fressin}, Francois and {Ge}, Jian and {Henning}, Thomas and {Holman}, Matthew J. and {Howard}, Andrew W. and {Ida}, Shigeru and {Jenkins}, Jon M. and {Jernigan}, Garrett and {Johnson}, John Asher and {Kaltenegger}, Lisa and {Kawai}, Nobuyuki and {Kjeldsen}, Hans and {Laughlin}, Gregory and {Levine}, Alan M. and {Lin}, Douglas and {Lissauer}, Jack J. and {MacQueen}, Phillip and {Marcy}, Geoffrey and {McCullough}, Peter R. and {Morton}, Timothy D. and {Narita}, Norio and {Paegert}, Martin and {Palle}, Enric and {Pepe}, Francesco and {Pepper}, Joshua and {Quirrenbach}, Andreas and {Rinehart}, Stephen A. and {Sasselov}, Dimitar and {Sato}, Bun'ei and {Seager}, Sara and {Sozzetti}, Alessandro and {Stassun}, Keivan G. and {Sullivan}, Peter and {Szentgyorgyi}, Andrew and {Torres}, Guillermo and {Udry}, Stephane and {Villasenor}, Joel},
        title = "{Transiting Exoplanet Survey Satellite (TESS)}",
      journal = {Journal of Astronomical Telescopes, Instruments, and Systems},
         year = 2015,
        month = jan,
       volume = {1},
          eid = {014003},
        pages = {014003},
          doi = {10.1117/1.JATIS.1.1.014003},
       adsurl = {https://ui.adsabs.harvard.edu/abs/2015JATIS...1a4003R}
}

@article{laughlin1998modification,
  title={The modification of planetary orbits in dense open clusters},
  author={Laughlin, Gregory and Adams, Fred C},
  journal={The Astrophysical Journal Letters},
  volume={508},
  number={2},
  pages={L171--L174},
  year={1998}
}

@article{breslau2019creating,
  title={Creating retrogradely orbiting planets by prograde stellar fly-bys},
  author={Breslau, Andreas and Pfalzner, Susanne},
  journal={Astronomy \& Astrophysics},
  volume={621},
  pages={A101},
  year={2019},
  publisher={EDP Sciences}
}

@ARTICLE{hen72,
       author = {{Henon}, M.},
        title = "{On the Simulation of Field Stars in Numerical Experiments}",
      journal = {\aap},
         year = 1972,
        month = jul,
       volume = {19},
        pages = {488},
       adsurl = {https://ui.adsabs.harvard.edu/abs/1972A&A....19..488H}
}

@ARTICLE{vercrefor09,
       author = {{Veras}, Dimitri and {Crepp}, Justin R. and {Ford}, Eric B.},
        title = "{Formation, Survival, and Detectability of Planets Beyond 100 AU}",
      journal = {\apj},
         year = 2009,
        month = may,
       volume = {696},
       number = {2},
        pages = {1600-1611},
          doi = {10.1088/0004-637X/696/2/1600},
archivePrefix = {arXiv},
       eprint = {0902.2779},
 primaryClass = {astro-ph.EP},
       adsurl = {https://ui.adsabs.harvard.edu/abs/2009ApJ...696.1600V}
}

@article{charalambous2025breaking,
  title={Breaking long-period resonance chains with stellar flybys},
  author={Charalambous, C and Cuello, N and Petrovich, C},
  journal={Astronomy \& Astrophysics},
  volume={696},
  pages={A175},
  year={2025},
  publisher={EDP Sciences}
}

@ARTICLE{col24,
       author = {{Coleman}, Gavin A.~L.},
        title = "{On the properties of free-floating planets originating in circumbinary planetary systems}",
      journal = {\mnras},
         year = 2024,
        month = may,
       volume = {530},
       number = {1},
        pages = {630-644},
          doi = {10.1093/mnras/stae903},
archivePrefix = {arXiv},
       eprint = {2403.18481},
 primaryClass = {astro-ph.EP},
       adsurl = {https://ui.adsabs.harvard.edu/abs/2024MNRAS.530..630C}
}

@ARTICLE{colder25,
       author = {{Coleman}, Gavin A.~L. and {DeRocco}, William},
        title = "{Predicting the Galactic population of free-floating planets from realistic initial conditions}",
      journal = {\mnras},
         year = 2025,
        month = mar,
       volume = {537},
       number = {3},
        pages = {2303-2312},
          doi = {10.1093/mnras/staf138},
archivePrefix = {arXiv},
       eprint = {2407.05992},
 primaryClass = {astro-ph.EP},
       adsurl = {https://ui.adsabs.harvard.edu/abs/2025MNRAS.537.2303C}
}

@ARTICLE{sitnes20,
       author = {{Sit}, Tawny and {Ness}, M.~K.},
        title = "{The Age Distribution of Stars in the Milky Way Bulge}",
      journal = {\apj},
         year = 2020,
        month = sep,
       volume = {900},
       number = {1},
          eid = {4},
        pages = {4},
          doi = {10.3847/1538-4357/ab9ff6},
archivePrefix = {arXiv},
       eprint = {2006.01158},
 primaryClass = {astro-ph.SR},
       adsurl = {https://ui.adsabs.harvard.edu/abs/2020ApJ...900....4S}
}

@ARTICLE{haszasfeu20,
       author = {{Hasselquist}, Sten and {Zasowski}, Gail and {Feuillet}, Diane K. and {Schultheis}, Mathias and {Nataf}, David M. and {Anguiano}, Borja and {Beaton}, Rachael L. and {Beers}, Timothy C. and {Cohen}, Roger E. and {Cunha}, Katia and {Fern{\'a}ndez-Trincado}, Jos{\'e} G. and {Garc{\'\i}a-Hern{\'a}ndez}, D.~A. and {Geisler}, Doug and {Holtzman}, Jon A. and {Johnson}, Jennifer and {Lane}, Richard R. and {Majewski}, Steven R. and {Moni Bidin}, Christian and {Nitschelm}, Christian and {Roman-Lopes}, Alexandre and {Schiavon}, Ricardo and {Smith}, Verne V. and {Sobeck}, Jennifer},
        title = "{Exploring the Stellar Age Distribution of the Milky Way Bulge Using APOGEE}",
      journal = {\apj},
         year = 2020,
        month = oct,
       volume = {901},
       number = {2},
          eid = {109},
        pages = {109},
          doi = {10.3847/1538-4357/abaeee},
archivePrefix = {arXiv},
       eprint = {2008.03603},
 primaryClass = {astro-ph.GA},
       adsurl = {https://ui.adsabs.harvard.edu/abs/2020ApJ...901..109H}
}

@ARTICLE{valzocmuc18,
       author = {{Valenti}, E. and {Zoccali}, M. and {Mucciarelli}, A. and {Gonzalez}, O.~A. and {Surot}, F. and {Minniti}, D. and {Rejkuba}, M. and {Pasquini}, L. and {Fiorentino}, G. and {Bono}, G. and {Rich}, R.~M. and {Soto}, M.},
        title = "{The central velocity dispersion of the Milky Way bulge}",
      journal = {\aap},
         year = 2018,
        month = aug,
       volume = {616},
          eid = {A83},
        pages = {A83},
          doi = {10.1051/0004-6361/201832905},
archivePrefix = {arXiv},
       eprint = {1805.00275},
 primaryClass = {astro-ph.GA},
       adsurl = {https://ui.adsabs.harvard.edu/abs/2018A&A...616A..83V}
}

@ARTICLE{lusinofe16,
       author = {{Lu}, J.~R. and {Sinukoff}, E. and {Ofek}, E.~O. and {Udalski}, A. and {Kozlowski}, S.},
        title = "{A Search For Stellar-mass Black Holes Via Astrometric Microlensing}",
      journal = {\apj},
         year = 2016,
        month = oct,
       volume = {830},
       number = {1},
          eid = {41},
        pages = {41},
          doi = {10.3847/0004-637X/830/1/41},
archivePrefix = {arXiv},
       eprint = {1607.08284},
 primaryClass = {astro-ph.SR},
       adsurl = {https://ui.adsabs.harvard.edu/abs/2016ApJ...830...41L}
}

@ARTICLE{lulamdaw19,
       author = {{Lu}, Jessica and {Lam}, Casey and {Dawson}, Will and {Gaudi}, B. Scott and {Golovich}, Nathan and {Medford}, Michael and {Abdurrahman}, Fatima and {Beaton}, Rachael L.},
        title = "{From Stars to Compact Objects: The Initial-Final Mass Relation}",
      journal = {\baas},
         year = 2019,
        month = may,
       volume = {51},
       number = {3},
          eid = {365},
        pages = {365},
          doi = {10.48550/arXiv.1904.01773},
archivePrefix = {arXiv},
       eprint = {1904.01773},
 primaryClass = {astro-ph.SR},
       adsurl = {https://ui.adsabs.harvard.edu/abs/2019BAAS...51c.365L}
}

@ARTICLE{hegfry03,
       author = {{Heger}, A. and {Fryer}, C.~L. and {Woosley}, S.~E. and {Langer}, N. and {Hartmann}, D.~H.},
        title = "{How Massive Single Stars End Their Life}",
      journal = {\apj},
         year = 2003,
        month = jul,
       volume = {591},
       number = {1},
        pages = {288-300},
          doi = {10.1086/375341},
archivePrefix = {arXiv},
       eprint = {astro-ph/0212469},
 primaryClass = {astro-ph},
       adsurl = {https://ui.adsabs.harvard.edu/abs/2003ApJ...591..288H}
}

@article{mctkipjoh20,
  title={8 in 10 Stars in the Milky Way Bulge experience stellar encounters within 1000 AU in a gigayear},
  author={{McTier}, Moiya AS and {Kipping}, David M and {Johnston}, Kathryn},
  journal={Monthly Notices of the Royal Astronomical Society},
  volume={495},
  number={2},
  pages={2105--2111},
  year={2020},
  publisher={Oxford University Press}
}

@article{penny2019predictions,
  title={Predictions of the WFIRST microlensing survey. I. Bound planet detection rates},
  author={Penny, Matthew T and Scott Gaudi, B and Kerins, Eamonn and Rattenbury, Nicholas J and Mao, Shude and Robin, Annie C and Calchi Novati, Sebastiano},
  journal={The Astrophysical Journal Supplement Series},
  volume={241},
  number={1},
  pages={3},
  year={2019},
  publisher={The American Astronomical Society}
}

@article{veras2013exoplanets,
  title={Exoplanets beyond the Solar neighbourhood: Galactic tidal perturbations},
  author={Veras, Dimitri and Evans, N Wyn},
  journal={Monthly Notices of the Royal Astronomical Society},
  volume={430},
  number={1},
  pages={403--415},
  year={2013},
  publisher={Oxford University Press}
}

@article{sumi2023free,
  title={Free-floating planet mass function from MOA-II 9 yr survey toward the galactic bulge},
  author={Sumi, Takahiro and Koshimoto, Naoki and Bennett, David P and Rattenbury, Nicholas J and Abe, Fumio and Barry, Richard and Bhattacharya, Aparna and Bond, Ian A and Fujii, Hirosane and Fukui, Akihiko and others},
  journal={The Astronomical Journal},
  volume={166},
  number={3},
  pages={108},
  year={2023},
  publisher={The American Astronomical Society}
}

@article{van2019survivability,
  title={Survivability of planetary systems in young and dense star clusters},
  author={van Elteren, Arjen and Portegies Zwart, Simon and Pelupessy, Inti and Cai, Maxwell X and McMillan, SLW},
  journal={Astronomy \& Astrophysics},
  volume={624},
  pages={A120},
  year={2019},
  publisher={EDP Sciences}
}

@article{wang2022hot,
  title={Hot Jupiter formation in dense clusters: secular chaos in multiplanetary systems},
  author={Wang, Yi-Han and Perna, Rosalba and Leigh, Nathan WC and Shara, Michael M},
  journal={Monthly Notices of the Royal Astronomical Society},
  volume={509},
  number={4},
  pages={5253--5264},
  year={2022},
  publisher={Oxford University Press}
}

@article{rodet2021correlation,
  title={On the correlation between hot Jupiters and stellar clustering: High-eccentricity migration induced by stellar flybys},
  author={Rodet, Laetitia and Su, Yubo and Lai, Dong},
  journal={The Astrophysical Journal},
  volume={913},
  number={2},
  pages={104},
  year={2021},
  publisher={The American Astronomical Society}
}

@article{raymond2024future,
  title={Future trajectories of the Solar System: dynamical simulations of stellar encounters within 100 au},
  author={Raymond, Sean N and Kaib, Nathan A and Selsis, Franck and Bouy, Herve},
  journal={Monthly Notices of the Royal Astronomical Society},
  volume={527},
  number={3},
  pages={6126--6138},
  year={2024},
  publisher={Oxford University Press}
}

@ARTICLE{yu2024free,
       author = {{Yu}, Fangyuan and {Lai}, Dong},
        title = "{Free-floating Planets, Survivor Planets, Captured Planets, and Binary Planets from Stellar Flybys}",
      journal = {\apj},
         year = 2024,
        month = jul,
       volume = {970},
       number = {1},
          eid = {97},
        pages = {97},
          doi = {10.3847/1538-4357/ad4f81},
archivePrefix = {arXiv},
       eprint = {2403.07224},
 primaryClass = {astro-ph.EP},
       adsurl = {https://ui.adsabs.harvard.edu/abs/2024ApJ...970...97Y}
}

@article{veras2014great,
  title={The great escape--III. Placing post-main-sequence evolution of planetary and binary systems in a Galactic context},
  author={Veras, Dimitri and Evans, N Wyn and Wyatt, Mark C and Tout, Christopher A},
  journal={Monthly Notices of the Royal Astronomical Society},
  volume={437},
  number={2},
  pages={1127--1140},
  year={2014},
  publisher={Oxford University Press}
}

@article{han2005microlensing,
  title={Microlensing detection and characterization of wide-separation planets},
  author={Han, Cheongho and Gaudi, B Scott and An, Jin H and Gould, Andrew},
  journal={The Astrophysical Journal},
  volume={618},
  number={2},
  pages={962--972},
  year={2005}
}

@article{mroz2020free,
  title={A free-floating or wide-orbit planet in the microlensing event OGLE-2019-BLG-0551},
  author={Mr{\'o}z, Przemek and Poleski, Rados{\l}aw and Han, Cheongho and Udalski, Andrzej and Gould, Andrew and Szyma{\'n}ski, Micha{\l} K and Soszy{\'n}ski, Igor and Pietrukowicz, Pawe{\l} and Koz{\l}owski, Szymon and Skowron, Jan and others},
  journal={The Astronomical Journal},
  volume={159},
  number={6},
  pages={262},
  year={2020},
  publisher={The American Astronomical Society}
}

@article{johnson2020predictions,
  title={Predictions of the nancy grace roman space telescope galactic exoplanet survey. II. Free-floating planet detection rates},
  author={Johnson, Samson A and Penny, Matthew and Gaudi, B Scott and Kerins, Eamonn and Rattenbury, Nicholas J and Robin, Annie C and Calchi Novati, Sebastiano and Henderson, Calen B},
  journal={The Astronomical Journal},
  volume={160},
  number={3},
  pages={123},
  year={2020},
  publisher={The American Astronomical Society}
}

@inproceedings{schlieder2024survey,
  title={Survey science with the Nancy Grace Roman Space Telescope wide field instrument},
  author={Schlieder, Joshua E and Barclay, Thomas and Barnes, Amethyst and Bray, Evan and Choi, Ami and Cromey, Benjamin and Delker, Thomas and Finch, Timothy and Frater, Eric H and Hill, Robert J and others},
  booktitle={Space Telescopes and Instrumentation 2024: Optical, Infrared, and Millimeter Wave},
  volume={13092},
  pages={196--221},
  year={2024},
  organization={SPIE}
}

@ARTICLE{2025arXiv250510574O,
       author = {{Observations Time Allocation Committee}, Roman and {Community Survey Definition Committees}, Core},
        title = "{Roman Observations Time Allocation Committee: Final Report and Recommendations}",
      journal = {arXiv e-prints},
         year = 2025,
        month = may,
          eid = {arXiv:2505.10574},
        pages = {arXiv:2505.10574},
          doi = {10.48550/arXiv.2505.10574},
archivePrefix = {arXiv},
       eprint = {2505.10574},
 primaryClass = {astro-ph.IM},
       adsurl = {https://ui.adsabs.harvard.edu/abs/2025arXiv250510574O}
}

\appendix

\section{Independence of results on planet mass}\label{sec:mass-indep}

\begin{figure*}
    \centering
    \includegraphics[width=\linewidth]{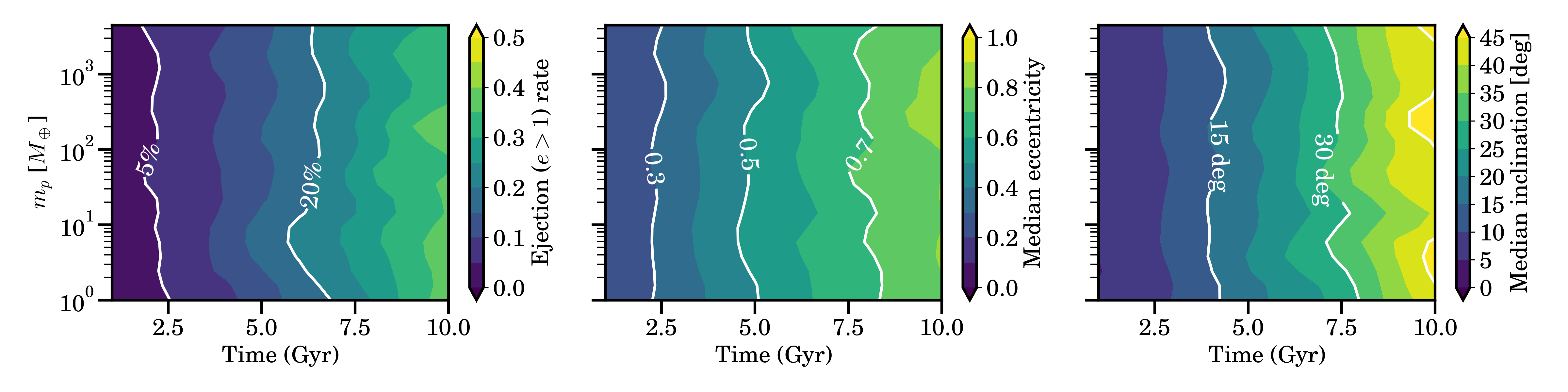}
    \caption{Distribution of ejection rate (left panel), median final eccentricity (middle panel), and median final inclination (right panel) as a function of integration time and planet mass. Here, the planets are initialized with $a_0=300\ \text{au}$ around solar mass hosts in a stellar environment of density $n_\star=19\ \text{stars}/\text{pc}^3$. The maximum perturber mass is $M_{\max}=10\ M_\odot$.}
    \label{fig:mass-dependence}
\end{figure*}
Here, we justify our choice of only considering planets of mass $m_p=10\ M_\oplus$ in our main simulation samples. Previous work has shown that the perturbation strength is independent of the mass of the bound body for planetary and brown dwarf companions \citep{van2019survivability}. We confirm this expectation numerically by carrying out additional integrations with $m_p \in [1\ M_\oplus, 15\ M_J]$. The results of these simulations are shown in \Cref{fig:mass-dependence}. As expected, the ejection rate, median eccentricity, and median inclination of planets which suffered repeated flybys is approximately independent of $m_p$.

\section{Choice of encounter sphere radius}\label{sec:denc}

\begin{figure}
    \centering
    \includegraphics[width=0.6\linewidth]{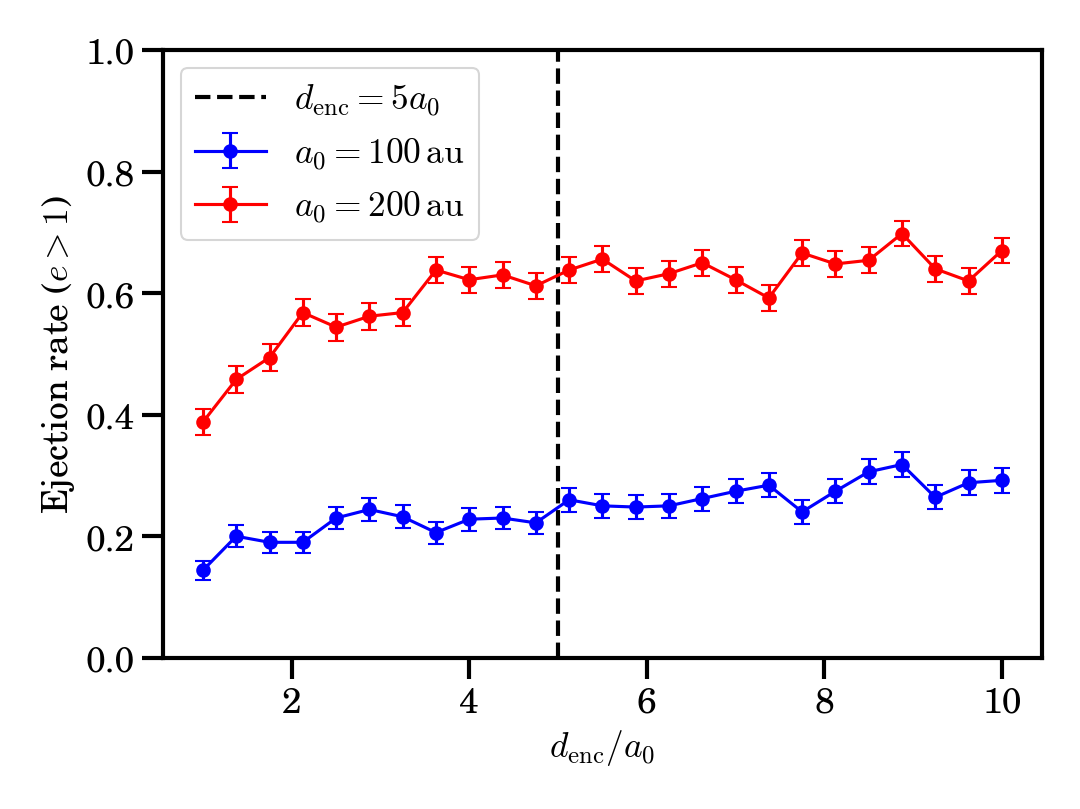}
    \caption{Ejection rate vs. $d_{\rm enc}/a_0$ for planets subject to the stellar environment with fiducial density ($n_\star{=}19\ \rm pc^{-3}$) and $M_{\rm max}{=}10\ M_\odot$. The planets orbit M star hosts ($M_\star{=}0.3 \ M_\odot$) at $a_0{=}\{100,200\}\ \rm au$. The error bars indicate the standard error for binomial distributions. For $d_{\rm enc}{\gtrsim} 5\ \rm au$, the ejection rates become independent of $d_{\rm enc}$.}
    \label{fig:renc-choice}
\end{figure}

Here, we justify our choice of encounter radius $d_{\rm enc} = 5a_0$ for planets at an initial semi-major axis $a_0$. For large values of $d_{\rm enc}/a_0$, the system will undergo a large number of flybys which do not penetrate the planet's orbit and thus amount to a weak perturbation. However, for values of $d_{\rm enc}/a_0$ which are too small, we risk missing flybys which may perturb the system. As such, we plot the ejection rate as a function of $d_{\rm enc}/a_0$ for $a_0 =100\, \rm au$ and $200\, \rm au$ in \Cref{fig:renc-choice}. We find that for $d_{\rm enc}/a_0 \sim 5$, the ejection rate plateaus, indicating that the flybys outside of $5a_0$ do not meaningfully impact the ejection dynamics. This justifies our choice to truncate the flyby distribution to only those which pass within $5a_0$ of the host star.

\end{document}